\documentclass[twocolumn,american,aps,prb,superscriptaddress]{revtex4-2}
\usepackage[T1]{fontenc}
\usepackage[utf8]{inputenc}
\usepackage{amsmath}
\usepackage{amssymb}
\usepackage{graphicx}

\makeatletter
\usepackage{babel}

\makeatother

\usepackage{babel}
\begin{document}
\title{Time-resolved ARPES in pumped excitonic systems:\\
Floquet physics induced by excitonic fields}
\author{Amir Eskandari-asl}
\affiliation{Dipartimento di Fisica ``E.R. Caianiello'', Università degli Studi
di Salerno, I-84084 Fisciano (SA), Italy}
\author{Adolfo Avella}
\affiliation{Dipartimento di Fisica ``E.R. Caianiello'', Università degli Studi
di Salerno, I-84084 Fisciano (SA), Italy}
\affiliation{CNR-SPIN, Unità di Salerno, I-84084 Fisciano (SA), Italy}
\affiliation{CNISM, Unità di Salerno, Università degli Studi di Salerno, I-84084
Fisciano (SA), Italy}
\begin{abstract}
We develop a theoretical framework based on the Dynamical Projective
Operatorial Approach (DPOA) to study the time- and angle-resolved
photoemission spectroscopy (TR-ARPES) of pumped excitonic systems.
Including Coulomb electron--electron interactions at the Hartree-Fock
(HF) level, our formalism captures the formation of excitonic bound
states under the application of pump pulses. Considering a prototypical
two-dimensional two-band semiconductor, we analyze the equilibrium
phase diagram, which shows the expected transition from a semiconducting
to an excitonic-insulator phase as the Coulomb interaction strength
or its range increase. Out of equilibrium, we find that when the pump
frequency is resonant with an excitonic mode, coherent oscillations
of the excitonic order parameter persist after the pump pulse subsides
and give rise to clear Floquet sidebands in the TR-ARPES spectrum.
These exciton-field-induced sidebands are distinct from those originating
from the pump laser field. We also identify band-resonance-induced
sidebands arising from residual coherences at momenta where the band
gap is resonant with the pump frequency. Finally, we analyze the local
Coulomb interaction limit. Our results corroborate recent experimental
observations of exciton-field-induced Floquet-like sidebands and establish
DPOA as an efficient and accurate method for simulating ultrafast
phenomena in interacting electron systems.
\end{abstract}
\maketitle

\section{Introduction}

The development of ultrafast pump--probe spectroscopy has opened
a new frontier in exploring the dynamics of condensed matter systems
on femtosecond and even sub-femtosecond time scales \citep{brabec2000intense,krausz2009attosecond,krausz2014attosecond,calegari2016advances,gandolfi2017emergent,borrego2022attosecond,inzani2023field,inzani2023photoinduced,inzani2026attosecond}.
By monitoring the temporal evolution of photo-excited carriers upon
perturbation by ultrashort and intense electromagnetic pulses, these
techniques provide direct access to the microscopic mechanisms governing
electronic, spin, and lattice degrees of freedom out of equilibrium
\citep{Zurch_17,doi:10.1063/1.4985056,PhysRevB.97.205202,perfetti2008femtosecond,schmitt2008transient}.
From a technological standpoint, such insight is essential for advancing
next-generation ultrafast optoelectronic and spintronic devices. From
a fundamental perspective, it enables probing pulse-induced symmetry
breaking, coherence, and relaxation processes in real time \citep{Zurch_17,PhysRevB.97.205202,perfetti2008femtosecond}.
One avenue to probe the response of a solid requires that the system
is excited by an intense pump pulse, typically in the infrared regime
with a duration ranging from a few to hundreds of femtoseconds, and
subsequently analyzed using a positively or negatively delayed probe
pulse. This analysis can measure either the transient change in optical
properties \citep{schultze2013controlling,stojchevska2014ultrafast,schultze2014attosecond,lucchini2016attosecond,mashiko2016petahertz,Borja:16,zurch2017ultrafast,Zurch_17,schlaepfer2018attosecond,Kaplan:19,geneaux2019,inzani2023field,inzani2023photoinduced,neufeld2023attosecond,inzani2026attosecond}
or the time-resolved angle-resolved photoemission spectroscopy (TR-ARPES)
signal \citep{schmitt2008transient,rohwer2011collapse,smallwood2012tracking,hellmann2012time,papalazarou2012coherent,wang2013observation,johannsen2013direct,rameau2016energy,reimann2018subcycle}.

Angle-resolved photoemission spectroscopy (ARPES) probes the electronic
band structure of materials by analyzing the energy and momentum of
photoejected electrons \citep{smith1975angular,himpsel1978experimental,kampf1990spectral,smith1991electronic,damascelli2003angle,hufner2013photoelectron,sobota2021angle}.
In pump-probe setups, TR-ARPES extends this capability to out-of-equilibrium
systems by measuring the signal as a function of the pump-probe delay
\citep{sobota2021angle,bovensiepen2012elementary,smallwood2016ultrafast,zhou2018new,boschini2024time}.
This technique reveals dynamical processes \citep{sobota2021angle}
of fundamental importance for understanding and engineering materials,
providing access to bands above the Fermi energy beyond what equilibrium
measurements can achieve \citep{weinelt2002time,echenique2004decay,sobota2013direct}.
TR-ARPES can also probe the dressing of bands and the emergence of
pump-induced Floquet sidebands \citep{wang2013observation,broers2022detecting},
as well as complex effects such as the perturbation of ordered states
\citep{schmitt2008transient,rohwer2011collapse,smallwood2012tracking}
and the dynamical excitation of collective modes \citep{papalazarou2012coherent,avigo2013coherent,yang2015inequivalence}.

Understanding these phenomena requires theoretical modeling and numerical
simulation. The standard approach, time-dependent density functional
theory (TD-DFT) \citep{de2013inside,de2016monitoring,wopperer2017efficient,de2017first,pemmaraju2018velocity,tancogne2018atomic,schlaepfer2018attosecond,neufeld2023attosecond,de2018real},
is computationally expensive and often limits physical insight \citep{armstrong2021dialogue}.
Model-Hamiltonian approaches, which typically use parameters derived
from equilibrium DFT calculations \citep{schuler2021gauge}, offer
a more flexible alternative, allowing controlled investigation of
individual terms and their interplay, but they often suffer from oversimplification
\citep{armstrong2021dialogue}. To address these challenges, we recently
developed the Dynamical Projective Operatorial Approach (DPOA) \citep{inzani2023field,eskandari2024time,eskandari2024out,eskandari2024generalized,eskandari2025dynamical},
an efficient operator-based formalism for studying ultrafast phenomena
in realistic multi-band systems. DPOA enables real-time evolution
of composite operators \citep{mancini2004} under intense pump fields
while maintaining access to all microscopic observables, including
excitation populations \citep{inzani2023field,eskandari2024time,eskandari2025controlling,eskandari2026controlling}
and multi-time response functions such as TR-ARPES \citep{eskandari2024time}
and transient optical properties \citep{eskandari2024generalized,eskandari2026magneto,inzani2026attosecond}.

On the other hand, the pursuit of non-trivial phases of matter through
optical driving has been a persistent objective in condensed matter
physics \citep{Lloyd-Hughes_2021,delatorre2021colloquium,giustino2020quantum}.
The oscillatory nature of the pump field itself provides a natural
connection to the concept of Floquet engineering, where a time-periodic
external drive is used to reshape the electronic band structure of
a material \citep{basov2017towards,hubener2017creating,claassen2016all}.
By hybridizing electronic states with the photons of an intense laser
field, one can create Floquet-Bloch states, leading to the emergence
of novel phases of matter such as Floquet topological insulators or
Weyl semimetals \citep{wang2013observation,mahmood2016selective,mciver2020light,aeschlimann2021survival}.
In essence, pump-probe experiments can be viewed as a realization
of Floquet physics in a pulsed, rather than continuous-wave, regime,
where the finite duration of the pump pulse introduces additional
temporal structure while preserving the fundamental mechanism of light-induced
band hybridization \citep{eskandari2024time}. Despite the immense
theoretical potential, the experimental realization of Floquet physics
in solids faces significant hurdles; consequently, the observation
of clear Floquet gaps and band hybridization has remained limited
to a few specific systems and conditions \citep{keunecke2020electromagnetic,hubener2018phonon}.

Recent proposals have turned to bosonic excitations beyond photons
as a means of generating time-periodic driving. Examples include phonon-driven
Floquet matter \citep{hubener2018phonon}, Floquet superconductivity
\citep{zhang2021anomalous}, and exciton-driven Floquet physics \citep{chan2023giant,perfetto2020floquet}.
In particular, the Coulomb interaction in low-dimensional semiconductors
leads to the formation of excitons---bound states of an electron
and a hole---with noticeable binding energies and transition dipole
moments, which can provide a large internal field. In a recent work,
Pareek \emph{et al.}~\citep{pareek2026driving} pursued this direction.
In a monolayer semiconductor, an exciton creates an oscillation of
the self-energy: the excitonic field. Such a field can drive Floquet-like
hybridization. Using TR-ARPES, they directly observed the modification
of the electronic bands in a monolayer transition metal dichalcogenide.
Upon resonant excitation of an excitonic mode, the valence band develops
a clear sideband, as well as the so called ``Mexican-hat'' dispersion
shape.

Several theoretical studies of TR-ARPES signals in pumped excitonic
systems have been performed \citep{perfetto2016first,rustagi2018photoemission,ohnishi2018direct,christiansen2019theory}.
A recently developed approach is the \emph{ab initio} time-dependent
GW (TD-aGW) approach \citep{chan2023giant,pareek2026driving}. In
this approach, the dynamical equations of the single-particle density
matrix are solved and then used to compute the two-time Green's functions,
which in turn enter the calculation of the TR-ARPES response. Such
an approach is indeed quite resource consuming, as it involves not
only solving the dynamical equations but also computing the double
integrals required for the TR-ARPES signal \citep{freericks2009theoretical}.
As we will demonstrate in this work, one can calculate the TR-ARPES
signal of pumped excitonic systems using DPOA, which has a noticeably
low computational cost.

Along this line, in this work, we extend the DPOA framework to investigate
pumped interacting systems. We develop a theoretical description that
incorporates Coulomb electron--electron interaction at the Hartree-Fock
(HF) level, allowing for the formation of excitonic bound states both
in equilibrium and under the influence of a pump pulse. We compute
the transient electronic structure and analyze the emergence of Floquet
sidebands in the TR-ARPES spectrum driven by the excitonic field.
Our approach captures the essential physics of exciton formation and
its dynamical evolution out of equilibrium. We find that when the
pump frequency matches an excitonic resonance, coherent oscillations
of the excitonic order parameter persist after the pump pulse and
give rise to Floquet sidebands in the TR-ARPES spectrum, which confirms
the findings of Ref. \citep{pareek2026driving}. These Floquet sidebands
are distinct from those originating from light field, and provide
a direct fingerprint of excitonic coherence. Moreover, our formalism
allows for a detailed analysis of the pump-induced features in the
presence of excitonic modes and more generally, Coulomb electron--electron
interaction. Our results demonstrate that TR-ARPES can serve as a
powerful tool to detect and characterize excitonic dynamics in pumped
semiconductors, and they establish DPOA as an efficient and accurate
method for simulating ultrafast phenomena in interacting electronic
systems.

The paper is organized as follows. In Sec.~\ref{sec:Theory}, we
present the theoretical formulation of the Coulomb electron--electron
interaction Hamiltonian, its HF treatment, the time evolution under
optical pumping, and the computation of the TR-ARPES signal within
the DPOA framework. In Sec.~\ref{sec:Numerical-studies}, we apply
our formalism to a prototypical two-dimensional two-band (valence-conduction)
semiconductor to investigate the two equilibrium phases (the semiconducting
and the excitonic-insulator phases) and the out-of-equilibrium TR-ARPES
response for both of them. We mainly consider a long-range Coulomb
electron-electron interaction, but we will also consider the special
case of the local Coulomb electron-electron interaction at the end
of the same section. Finally, Sec.~\ref{sec:Conclusions}, we summarize
our conclusions.

\section{Theory\label{sec:Theory}}

\subsection{Coulomb electron--electron interaction Hamiltonian}

The Hamiltonian governing a non-interacting lattice system can be
written as \citep{eskandari2024time}
\begin{equation}
H_{0}=\sum_{\mathbf{k}}c^{\dagger}\left(\mathbf{k}\right)\cdot\Xi^{0}\left(\mathbf{k}\right)\cdot c\left(\mathbf{k}\right),
\end{equation}
where $c\left(\mathbf{k}\right)$ is a vector whose components are
the annihilation operators, $c_{l}\left(\mathbf{k}\right)$, for an
electron with the set of quantum numbers (such as band, spin, etc.)
$l$ and with momentum $\mathbf{k}$, $\cdot$ is used for matrix
multiplication in such a space of quantum numbers (it will also be
used as scalar contraction in Cartesian space), and $\Xi^{0}\left(\mathbf{k}\right)$
is obtained by diagonalizing the hopping matrix expressed in the basis
of localized Wannier functions, $\tilde{T}_{\mathbf{k}}$ , via the
unitary transformation given by the matrix $\Omega_{\mathbf{k}}$,
\begin{align}
\Xi^{0}\left(\mathbf{k}\right)=\Omega^{\dagger}_{\mathbf{k}}\cdot\tilde{T}_{\mathbf{k}}\cdot\Omega_{\mathbf{k}},
\end{align}
where $\Xi^{0}_{l,l^{\prime}}\left(\mathbf{k}\right)=\delta_{l,l^{\prime}}\varepsilon^{0}_{l}\left(\mathbf{k}\right)$,
in which $\delta_{l,l^{\prime}}$ is the Kronecker delta and $\varepsilon^{0}_{l}\left(\mathbf{k}\right)$
represent the non-interacting band energies.

The Coulomb electron--electron interaction is taken into account
through the following density--density term 
\begin{equation}
H_{V}=\frac{1}{2}\sum_{\mathbf{i},\mathbf{i}'}\sum_{l,l^{\prime}}V_{l,l^{\prime}}\left(\mathbf{i}-\mathbf{i}'\right)n_{l}\left(\mathbf{i}\right)n_{l^{\prime}}\left(\mathbf{i}'\right),
\end{equation}
where $V_{l,l^{\prime}}\left(\mathbf{i}-\mathbf{i}'\right)$ is the
interaction potential matrix element, $\mathbf{i}$ and $\mathbf{i}'$
run over lattice sites, and $n_{l}\left(\mathbf{i}\right)=c^{\dagger}_{l}\left(\mathbf{i}\right)c_{l}\left(\mathbf{i}\right)$.
The real-space operators are linked to their momentum-space counterparts
through $c_{l}\left(\mathbf{i}\right)=\frac{1}{\sqrt{M}}\sum_{\mathbf{k}}\mathrm{e}^{-\mathrm{i}\mathbf{k}\cdot\mathbf{R_{i}}}c_{l}\left(\mathbf{k}\right)$,
with $M$ the total number of lattice sites and $\mathbf{R_{i}}$
the position of site $\mathbf{i}$. A direct Fourier transformation
yields 
\begin{multline}
H_{V}=\frac{1}{2M}\sum_{\mathbf{k}}\sum_{l,l^{\prime}}V_{l,l^{\prime}}\left(\mathbf{k}\right)\times\\
\times\sum_{\mathbf{q},\mathbf{q}'}c^{\dagger}_{l}\left(\mathbf{q}-\mathbf{k}\right)c_{l}\left(\mathbf{q}\right)c^{\dagger}_{l^{\prime}}\left(\mathbf{q}'+\mathbf{k}\right)c_{l^{\prime}}\left(\mathbf{q}'\right),
\end{multline}
in which we have introduced the Fourier transform of the interaction
potential matrix element, 
\begin{equation}
V_{l,l^{\prime}}\left(\mathbf{k}\right)=\sum_{\mathbf{i}}V_{l,l^{\prime}}\left(\mathbf{i}\right)\mathrm{e}^{-\mathrm{i}\mathbf{k}\cdot\mathbf{R_{i}}}.
\end{equation}

\subsection{Hartree--Fock approximation}

We treat the Coulomb interaction at the mean-field level by employing
the HF decoupling. Moreover, (i) we assume that the equilibrium expectation
values conserve crystal momentum, i.e., $\left\langle c^{\dagger}_{l}\left(\mathbf{q}\right)c_{l^{\prime}}\left(\mathbf{q}'\right)\right\rangle \propto\delta_{\mathbf{q},\mathbf{q}'}$,
and (ii) we remove the contributions that would be canceled by the
ionic positively-charged background (specifically, the terms proportional
to $V_{l,l^{\prime}}\left(\mathbf{k=0}\right)$). After these steps,
the HF-decoupled Coulomb interaction Hamiltonian acquires the form
\begin{align}
H_{V}\simeq H_{HF} & =\sum_{\mathbf{k}}c^{\dagger}\left(\mathbf{k}\right)\cdot\Xi^{HF}\left(\mathbf{k}\right)\cdot c\left(\mathbf{k}\right),
\end{align}
where 
\begin{equation}
\Xi^{HF}_{ll^{\prime}}\left(\mathbf{k}\right)=-\frac{1}{M}\sum_{\mathbf{q}}V_{l^{\prime}l}\left(\mathbf{k}-\mathbf{q}\right)\rho_{ll^{\prime}}\left(\mathbf{q}\right),\label{eq:S-HF}
\end{equation}
in which we have assumed $V_{ll^{\prime}}\left(-\mathbf{k}\right)=V_{l^{\prime}l}\left(\mathbf{k}\right)$.

Here $\rho_{ll^{\prime}}\left(\mathbf{k}\right)=\left\langle c^{\dagger}_{l^{\prime}}\left(\mathbf{k}\right)c_{l}\left(\mathbf{k}\right)\right\rangle $
is the single-particle density matrix (SPDM). It is worth noting that
although the original Coulomb interaction $H_{V}$ is repulsive, the
subtraction of the background term makes the effective HF Coulomb
interaction attractive, a crucial feature for the formation of excitonic
bound states. Note that $\Xi^{HF}\left(\mathbf{k}\right)$ is referred
to as the HF self-energy.

Because the HF Hamiltonian depends on the SPDM, a self-consistent
calculation is necessary to determine the equilibrium properties.
The iterative procedure adopted in this manuscript for such a self-consistent
loop is described in detail in Appendix~\ref{sec:SC_EQ}.

\subsection{Time evolution}

Once the self-consistent equilibrium solution has been obtained, we
examine the system's response to a pump pulse. The coupling to the
electric field is introduced within the standard minimal coupling
framework, but now the Hamiltonian becomes time-dependent through
both the pump field and the instantaneous HF potential. Consequently,
the full Hamiltonian can be cast in the form 
\begin{align}
H\left(t\right) & =\sum_{\mathbf{k}}c^{\dagger}\left(\mathbf{k}\right)\cdot\Xi\left(\mathbf{k},t\right)\cdot c\left(\mathbf{k}\right),\label{eq:H_t}
\end{align}
with 
\begin{equation}
\Xi\left(\mathbf{k},t\right)=\Xi^{0}\left(\mathbf{k}\right)+\Xi^{\text{pu}}\left(\mathbf{k},t\right)+\Xi^{HF}\left(\mathbf{k},t\right),\label{eq:S-full-t}
\end{equation}
where, following Refs.~\citep{eskandari2024time,eskandari2024generalized},
\begin{equation}
\Xi^{0}\left(\mathbf{k}\right)+\Xi^{\text{pu}}\left(\mathbf{k},t\right)=T_{\boldsymbol{k}}\left(t\right)+\mathrm{e}\boldsymbol{E}_{\text{pu}}\left(t\right)\cdot\boldsymbol{D}_{\boldsymbol{k}}\left(t\right),
\end{equation}
in which, $\boldsymbol{D}_{\boldsymbol{k}}$ corresponds to the dipole
moment, and for operator $O\in\left\{ T,\boldsymbol{D}\right\} $,
the time-dependent matrix is obtained through a generalized Peierls
substitution \citep{eskandari2024time}: 
\begin{align}
O_{\mathbf{k}}\left(t\right)=\Omega^{\dagger}_{\mathbf{k}}\cdot\tilde{O}_{\mathbf{k}+\frac{\mathrm{e}}{\hbar}\boldsymbol{A}_{\text{pu}}\left(t\right)}\cdot\Omega_{\mathbf{k}},
\end{align}
where $\tilde{O}$ refers to these matrices in the localized Wannier
basis, and $\boldsymbol{A}_{\text{pu}}\left(t\right)$ is the vector
potential of the pump field, related to the pump electric field by
$\boldsymbol{E}_{\text{pu}}\left(t\right)=-\partial_{t}\boldsymbol{A}_{\text{pu}}\left(t\right)$.
We operate within the dipole approximation, assuming the pump wavelength
is much larger than the lattice spacing.

The quadratic form of the time-dependent Hamiltonian, Eq.~\ref{eq:H_t},
suggests that the creation and annihilation operators are convenient
choices for the eigenoperators within DPOA \citep{eskandari2024time,eskandari2025dynamical}.
Therefore, in the Heisenberg picture we write $c\left(\mathbf{k},t\right)=P\left(\mathbf{k},t\right)\cdot c\left(\mathbf{k}\right)$,
where $c\left(\mathbf{k}\right)=c\left(\mathbf{k},t_{0}\right)$ corresponds
to the operators at an initial time $t_{0}\rightarrow-\infty$ , well
before the pump is applied, i.e., in equilibrium. The evolution matrix
$P\left(\mathbf{k},t\right)$ satisfies the equation of motion \citep{eskandari2024time}
\begin{equation}
\mathrm{i}\hbar\partial_{t}P\left(\mathbf{k},t\right)=\Xi\left(\mathbf{k},t\right)\cdot P\left(\mathbf{k},t\right),\label{eq:dyn_P}
\end{equation}
subject to the initial condition $P\left(\mathbf{k},t_{0}\right)=\mathbf{1}$.
At each instant of time, $\boldsymbol{\Xi}^{HF}\left(\mathbf{k},t\right)$
depends on the instantaneous SPDM, $\rho\left(\mathbf{k},t\right)$,
(through Eq.~\ref{eq:S-HF} by replacing $\rho\left(\boldsymbol{\text{q}}\right)\rightarrow\rho\left(\boldsymbol{\text{q}},t\right)$,
which itself is given by 
\begin{align}
\rho\left(\mathbf{k},t\right) & =P\left(\mathbf{k},t\right)\cdot\rho^{\text{eq}}\left(\mathbf{k}\right)\cdot P^{\dagger}\left(\mathbf{k},t\right),
\end{align}
where $\rho^{\text{eq}}\left(\mathbf{k}\right)=\rho\left(\mathbf{k},t_{0}\right)$
is the SPDM obtained from the equilibrium self-consistent calculation
(see Appendix~\ref{sec:SC_EQ}).

It is noteworthy that one must compute the dynamical version of Eq.~\ref{eq:S-HF}
(by substituting $\rho_{ll^{\prime}}\left(\boldsymbol{\text{q}}\right)\rightarrow\rho_{ll^{\prime}}\left(\boldsymbol{\text{q}},t\right)$)
at each time step and for each point on the numerical momentum grid.
As it is well known, such direct calculations are extremely resource
consuming. To make the computation feasible, we employ a Fast-Fourier-Transform
routine combined with the standard convolution approach, which significantly
reduces the numerical cost. For the time evolution of the dynamical
equations (in particular, Eq.~\ref{eq:dyn_P}) we use a fourth-order
Runge-Kutta approach.

\subsection{TR-ARPES signal}

To compute the TR-ARPES signal, we employ the formalism of non-equilibrium
Green's functions. We define the standard retarded and lesser Green's
functions as 
\begin{align}
 & G^{R}_{l,l^{\prime}}\left(\mathbf{k},t,t^{\prime}\right)=-\mathrm{i}\theta\left(t-t^{\prime}\right)\left\langle \left\{ c_{l}\left(\mathbf{k},t\right),c^{\dagger}_{l^{\prime}}\left(\mathbf{k},t^{\prime}\right)\right\} \right\rangle ,\\
 & G^{<}_{l,l^{\prime}}\left(\mathbf{k},t,t^{\prime}\right)=\mathrm{i}\left\langle c^{\dagger}_{l^{\prime}}\left(\mathbf{k},t^{\prime}\right)c_{l}\left(\mathbf{k},t\right)\right\rangle .
\end{align}
Within DPOA, one can demonstrate that these Green's functions can
be written as \citep{eskandari2024time}
\begin{equation}
G^{R}\left(\mathbf{k},t,t^{\prime}\right)=-\mathrm{i}\theta\left(t-t^{\prime}\right)P\left(\mathbf{k},t\right)\cdot P^{\dagger}\left(\mathbf{k},t^{\prime}\right),
\end{equation}
\begin{equation}
G^{<}\left(\mathbf{k},t,t^{\prime}\right)=\mathrm{i}P\left(\mathbf{k},t\right)\cdot\rho^{\text{eq}}\left(\mathbf{k}\right)\cdot P^{\dagger}\left(\mathbf{k},t^{\prime}\right).
\end{equation}

The experimentally accessible quantity is the photocurrent, which
we refer to as the electron TR-ARPES signal, or simply, just the TR-ARPES
signal. It is defined by convolving the lesser Green's function with
the probe pulse envelope \citep{freericks2009theoretical,eskandari2024time}:
\begin{multline}
I^{\text{e}}\left(\mathbf{k},\omega,t_{\mathrm{pr}}\right)=\frac{\tau_{\mathrm{pr}}}{\sqrt{8\pi\ln2}}\int^{+\infty}_{-\infty}dt\int^{+\infty}_{-\infty}dt^{\prime}S_{\mathrm{pr}}\left(t-t_{\mathrm{pr}}\right)\\
S_{\mathrm{pr}}\left(t^{\prime}-t_{\mathrm{pr}}\right)\Im\left[e^{\mathrm{i}\omega\left(t-t^{\prime}\right)}\text{Tr}\left[G^{<}\left(\mathbf{k},t,t^{\prime}\right)\right]\right],
\end{multline}
where $S_{\mathrm{pr}}(t)=\frac{2\sqrt{\ln2}}{\sqrt{\pi}\tau_{\mathrm{pr}}}e^{-4\ln2\,t^{2}/\tau^{2}_{\mathrm{pr}}}$
is the probe-pulse Gaussian envelope of full width at half maximum
(FWHM) $\tau_{\mathrm{pr}}$, and $t_{\mathrm{pr}}$ is the probe
delay. Note that this definition is a measure of the actual TR-ARPES
signal, but doesn't include the matrix elements coupling the free
electronic states in vacuum and the states in the material \citep{eskandari2024time}.
Accordingly, our $I^{\text{e}}$ has the dimensions of time, and the
overall coefficients have been chosen such that integrating over $\omega$
gives the total number of particles at each crystal momentum. $I^{\text{e}}$
reflects the occupied electronic states, but to map out the full band
structure (or the transient bands), one may instead use the retarded
Green's function, defining the quantity \citep{eskandari2024time}
\begin{multline}
I^{R}\left(\mathbf{k},\omega,t_{\mathrm{pr}}\right)=-\frac{\tau_{\mathrm{pr}}}{\sqrt{2\pi\ln2}}\int^{+\infty}_{-\infty}dt\int^{+\infty}_{-\infty}dt^{\prime}S_{\mathrm{pr}}\left(t-t_{\mathrm{pr}}\right)\\
S_{\mathrm{pr}}\left(t^{\prime}-t_{\mathrm{pr}}\right)\Im\left[e^{\mathrm{i}\omega\left(t-t^{\prime}\right)}\text{Tr}\left[G^{R}\left(\mathbf{k},t,t^{\prime}\right)\right]\right].
\end{multline}

Following analogous steps to those presented in Ref.~\citep{eskandari2024time},
these expressions can be recast in a compact form: 
\begin{align}
 & I^{\text{e}}\left(\mathbf{k},\omega,t_{\mathrm{pr}}\right)=\nonumber \\
 & \frac{\tau_{\mathrm{pr}}}{\sqrt{8\pi\ln2}}\text{Tr}\left[Q\left(\mathbf{k},\omega,t_{\mathrm{pr}}\right)\rho^{\text{eq}}\left(\mathbf{k}\right)Q^{\dagger}\left(\mathbf{k},\omega,t_{\mathrm{pr}}\right)\right],
\end{align}
\begin{equation}
I^{R}\left(\mathbf{k},\omega,t_{\mathrm{pr}}\right)=\frac{\tau_{\mathrm{pr}}}{\sqrt{8\pi\ln2}}\text{Tr}\left[Q\left(\mathbf{k},\omega,t_{\mathrm{pr}}\right)Q^{\dagger}\left(\mathbf{k},\omega,t_{\mathrm{pr}}\right)\right],
\end{equation}
where we have introduced the auxiliary matrix 
\begin{equation}
Q\left(\mathbf{k},\omega,t_{\mathrm{pr}}\right)=\int^{+\infty}_{-\infty}dt\,e^{\mathrm{i}\omega t}P\left(\mathbf{k},t\right)S_{\mathrm{pr}}\left(t-t_{\mathrm{pr}}\right).
\end{equation}
This procedure avoids directly computing many double integrals and
noticeably speeds up numerical calculations.

Experimentally, only the electron signal, $I^{\text{e}}$, is directly
measurable. Nevertheless, it is useful to define theoretically a hole
signal as 
\begin{equation}
I^{\text{h}}\left(\mathbf{k},\omega,t_{\mathrm{pr}}\right)=I^{R}\left(\mathbf{k},\omega,t_{\mathrm{pr}}\right)-I^{\text{e}}\left(\mathbf{k},\omega,t_{\mathrm{pr}}\right),
\end{equation}
which provides complementary information about the unoccupied states.

\section{Excitonic dynamics in a pumped two-band semiconductor\label{sec:Numerical-studies}}

\subsection{System and Equilibrium Phases}

\begin{figure}
\centering{}\includegraphics[width=8cm]{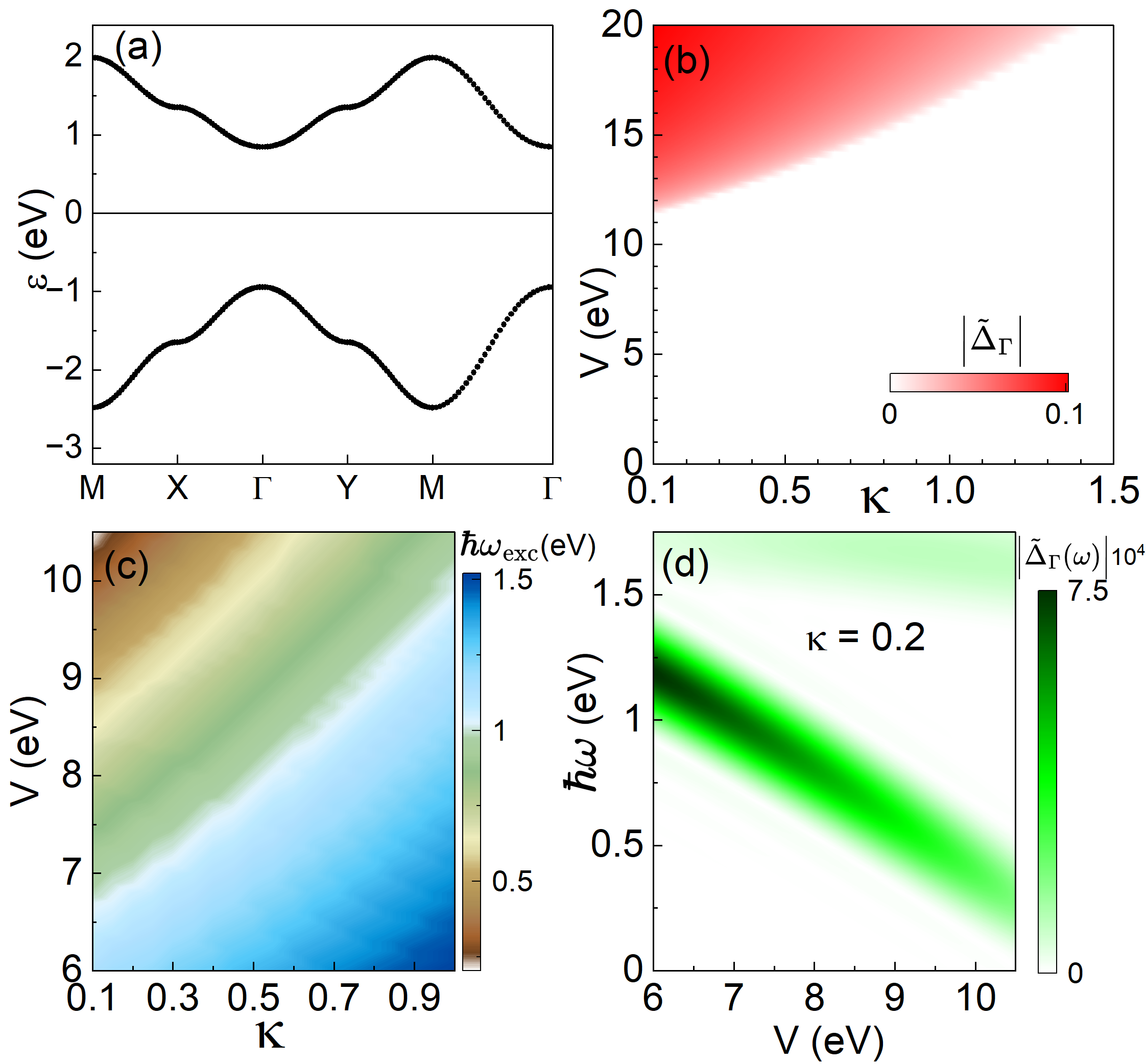} \caption{Equilibrium properties of the two-band model. (a) Non-interacting
band structure of the two-band model along a high-symmetry path in
the Brillouin zone. (b) Equilibrium normalized off-diagonal HF self-energy
at the $\Gamma$ point, $\tilde{\Delta}_{\Gamma}$, as a function
of the Coulomb interaction strength $V$ and the screening factor
$\kappa$. The boundary between the semiconducting phase ($\tilde{\Delta}_{\Gamma}=0$)
and the excitonic insulator phase ($\tilde{\Delta}_{\Gamma}\protect\neq0$)
is visible. (c) Main equilibrium excitonic energy, $\hbar\omega_{\mathrm{exc}}$,
in the semiconducting phase as a function of $V$ and $\kappa$. (d)
Fourier-transformed excitonic field, $\tilde{\Delta}_{\Gamma}(\omega)$,
as a function of $V$ and $\hbar\omega$ for a fixed screening factor
$\kappa=0.2$, showing the main excitonic mode and a weaker higher-energy
mode. \label{fig:system_eq_pd_wexc}}
\end{figure}

In this section, we apply our formalism to a prototypical two-dimensional
(2D) two-band (valence-conduction) model that mimics a semiconductor.
The lattice constant is denoted by $a$, and for simplicity we consider
spinless electrons. In the localized Wannier basis, the hopping matrix
elements are specified as follows. The onsite energies are $\tilde{T}_{\boldsymbol{R}=0,1,1}=-1.65\,\text{eV}$
and $\tilde{T}_{\boldsymbol{R}=0,2,2}=1.35\,\text{eV}$ for the two
Wannier states indexed by 1 and 2. The nearest-neighbor diagonal hoppings,
are $\tilde{T}^{\text{hop}}_{\mathbf{R}=\mathbf{a},1,1}=0.2\,\text{eV}$
and $\tilde{T}^{\text{hop}}_{\mathbf{R}=\mathbf{a},2,2}=-0.15\,\text{eV}$,
while the off-diagonal nearest-neighbor hopping is $\tilde{T}_{\mathbf{R}=\mathbf{a},1,2}=\tilde{T}_{\mathbf{R}=\mathbf{a},2,1}=-0.1\,\text{eV}$.
Here, $\tilde{T}_{\boldsymbol{R},\nu,\nu'}$ denotes the hopping matrix
element between Wannier states $\nu$ and $\nu'$ centered on lattice
sites separated by the vector $\boldsymbol{R}$, and $\mathbf{a}\in\{\pm a\hat{x},\pm a\hat{y}\}$
runs over all nearest-neighbor displacement vectors. The Brillouin
zone (BZ) is sampled by a uniform grid of $M=64\times64$ $\boldsymbol{k}$-points
including the $\Gamma$ point. Fourier transforming $\tilde{T}_{\boldsymbol{R}}$
yields the momentum-space hopping matrix $\tilde{T}_{\boldsymbol{k}}$
as described in Refs.~\citep{eskandari2024time,eskandari2024generalized}.
All calculations are performed at zero temperature, and the chemical
potential is set within the band gap. In Fig.~\ref{fig:system_eq_pd_wexc}(a),
we show the non-interacting bands of this system in equilibrium along
a high-symmetry path in the BZ.

If we keep the model perfectly centrosymmetric, in the semiconducting
phase, the excitons would remain \emph{dark}, because, for instance,
the first-order light-matter coupling at $\Gamma$ would vanish. To
break this symmetry, and make the excitons \emph{bright}, we introduce
a local dipole moment: $\tilde{\boldsymbol{D}}_{\boldsymbol{R}=0,1,2}=\tilde{\boldsymbol{D}}_{\boldsymbol{R}=0,2,1}=D\hat{x}$
with $D=0.1$\,nm. It is worth noticing that a similar effect could
have been achieved also by applying a left-right asymmetry in the
hopping matrix elements, or considering a model like SSH where the
selection rules allow for the non-zero coupling elements at $\Gamma$.

The expression of the Coulomb electron--electron interaction potential,
$V\left(\mathbf{k}\right)$, has been obtained by solving the discrete
Poisson equation on a three-dimensional (3D) lattice \citep{PhysRevB.54.12443,PhysRevLett.125.257002,PhysRevLett.129.047001,spalek2022superconductivity},
embedding our two-dimensional one. This latter resides at $z=0$ within
the embedding 3D lattice and its actual expression for $V\left(\mathbf{k}\right)$
has been computed integrating out $k_{z}$ from the corresponding
3D expression \citep{PhysRevB.54.12443}. Accordingly, we have
\begin{equation}
V\left(\mathbf{k}\right)=\frac{V}{\sqrt{\left(\tilde{k}^{2}+\kappa^{2}+2\right)^{2}-4}},\label{eq:V_k_LR}
\end{equation}
where $\tilde{k}^{2}=2\left(2-\cos k_{x}a-\cos k_{y}a\right)$ and
$\kappa$ controls a Yukawa-like screening mimicking the effects on
the dielectric function of a small, but finite, carrier mobility.
Here, we have assumed that the 3D lattice is cubic ($a_{x}=a_{y}=a_{z}=a$)
and that the system exhibits an isotropic dielectric function. This
long-range interaction will be considered for most of the results
we analyze. At the end of the results section, we will also consider
the local interaction case and, therefore, a momentum independent
Coulomb electron--electron interaction potential.

We choose the Coulomb interaction to act only between the conduction
and the valence bands (the CBs and the VBs, respectively), because
our main focus in this study is the electron-hole bound states. Thus,
the only non-zero matrix elements are $V_{v,c}\left(\mathbf{k}\right)=V_{c,v}\left(\mathbf{k}\right)=V\left(\mathbf{k}\right)$,
where $v$ and $c$ refer to the VB and the CB, respectively, and
$V$ is dubbed the Coulomb interaction strength. As such, the only
relevant term in the HF self-energy is the off-diagonal one. To analyze
the excitonic field, we consider the value of the off-diagonal HF
self-energy divided by the Coulomb interaction strength $V$ and denote
it, in equilibrium, by 
\begin{equation}
\tilde{\Delta}_{\mathbf{k}}=\frac{\Xi^{\mathrm{HF}}_{c,v}(\mathbf{k})}{V}.
\end{equation}
As a representative quantity that captures the excitonic order parameter
at the center of the BZ, we focus mainly on $\tilde{\Delta}_{\Gamma}=\tilde{\Delta}_{\mathbf{k}=\mathbf{0}}$.
Out of equilibrium, this quantity becomes a function of time and we
denote it by $\tilde{\Delta}_{\Gamma}\left(t\right)$. Performing
the Fourier transform with respect to time, we obtain it as a function
of the angular frequency, $\omega$, and denote it by $\tilde{\Delta}_{\Gamma}\left(\omega\right)$.

To characterize the equilibrium phases, we study $\tilde{\Delta}_{\Gamma}$
in equilibrium. In Fig.~\ref{fig:system_eq_pd_wexc}(b), we show
$\tilde{\Delta}_{\Gamma}$ as a function of the Coulomb interaction
strength $V$ and the screening factor $\kappa$. For each given $\kappa$,
there is a critical value $V$ below which $\tilde{\Delta}_{\Gamma}=0$
and the system has no excitonic bound state in its ground state; we
refer to this regime as the semiconducting phase. In this phase, the
ground state SPDM is the same as the one of the non-interacting system,
which makes the HF self-energy vanish, and hence the equilibrium bands
also coincide with the non-interacting ones. Above this critical $V$,
the equilibrium value of $\tilde{\Delta}_{\Gamma}$ becomes finite,
indicating the spontaneous formation of excitonic bound states in
the ground state. We will refer to such a phase as the excitonic insulator
phase. It is noteworthy that the critical $V$ increases as $\kappa$
increases. This is because a larger $\kappa$ means a more localized
(shorter-range) Coulomb interaction, and reducing the range of the
interaction decreases also its effective strength, thus requiring
a larger bare $V$ to reach the excitonic instability.

\subsection{Equilibrium Excitonic Frequency For the Semiconducting Phase}

In the semiconducting phase, there is no excitonic bound state in
the ground state. However, one may pump the system, and if the pump
frequency is in a specific range, excitonic bound states form. Such
frequency ranges determine what we call hereafter the excitonic frequencies.
In such a scenario, after the application of the pump pulse, the excitonic
self-energy oscillates at the excitonic frequency \citep{chan2023giant,pareek2026driving}.
This excitonic frequency depends on the pump-pulse parameters (amplitude
and duration) if the pump pulse is intense and long enough, but, for
weak pump intensities, it is an intrinsic property of the system,
to which we refer as the equilibrium excitonic frequency.

In order to determine the equilibrium excitonic frequency, we have
devised the following protocol. We apply a very weak impulsive pump
pulse with a Dirac-like delta function shape in time (i.e., a very
narrow Gaussian), resulting in a wide frequency content. Then, we
compute $\tilde{\Delta}_{\Gamma}(\omega)$ and find the frequency
$\omega_{\mathrm{exc}}$ for which $\left|\tilde{\Delta}_{\Gamma}(\omega_{\mathrm{exc}})\right|$
is maximum among all positive $\hbar\omega$ smaller than the energy
band gap at $\Gamma$. The quantity $\hbar\omega_{\mathrm{exc}}$
is the main equilibrium excitonic energy of the system. It is worth
noting that, in general, $\left|\tilde{\Delta}_{\Gamma}(\omega)\right|$
can have several local maxima, all of which would determine the equilibrium
excitonic frequencies corresponding to different excitonic states.

To numerically implement such an impulsive pump pulse, we consider
the following vector potential 
\begin{equation}
A_{\mathrm{imp}}(t)=A_{0,\mathrm{imp}}\,e^{-(4\ln2)t^{2}/\tau^{2}_{\mathrm{imp}}},
\end{equation}
with $\tau_{\mathrm{imp}}=0.03\,\text{fs}$ and $A_{0,\mathrm{imp}}=0.05\,\text{V}\cdot\text{fs}/\text{nm}$
(the robustness of the results with respect to reasonable variations
of these parameters has been verified). In Fig.~\ref{fig:system_eq_pd_wexc}(c),
we report the main equilibrium excitonic energy, $\hbar\omega_{\mathrm{exc}}$,
as a function of $V$ and $\kappa$ for ranges that fall into the
semiconducting phase. As it can be seen, by increasing the Coulomb
interaction strength $V$, or by making it more long-range (decreasing
$\kappa$), the main equilibrium excitonic energy decreases. This
behavior occurs because a stronger or more long-range interaction
leads to a more strongly bound electron--hole pair, which requires
less energy to be excited. Approaching the critical Coulomb interaction
parameters (the boundary of the semiconducting phase), the equilibrium
excitonic energy approaches zero, meaning that one can have excitonic
bound states with a negligibly small photon energy.

The full dependence of $\left|\tilde{\Delta}_{\Gamma}(\omega)\right|$
for the specific case of $\kappa=0.2$ is shown in Fig.~\ref{fig:system_eq_pd_wexc}(d)
as a function of $V$ and $\hbar\omega$. It is clear that increasing
$V$ reduces the excitonic energy, consistently with the trend observed
in panel (c). Moreover, we observe another weak excitonic mode at
a higher energy, which corresponds to another excitonic state. The
time interval for these Fourier transformations is from 2 to 32 fs.
Moreover, Hann windowing has been applied to all Fourier transformation
of the manuscript to reduce noise.

\subsection{Out-of-Equilibrium: Semiconducting Phase}

To study the out-of-equilibrium dynamics, we apply a linearly polarized
pump pulse with vector potential $\boldsymbol{A}_{\mathrm{pu}}(t)=A_{\mathrm{pu}}(t)\hat{x}$,
where 
\begin{equation}
A_{\mathrm{pu}}(t)=A_{0}\,e^{-(4\ln2)t^{2}/\tau^{2}_{\mathrm{pu}}}\sin(\omega_{\mathrm{pu}}t).
\end{equation}
The pulse has a Gaussian envelope with full width at half maximum
(FWHM) $\tau_{\mathrm{pu}}=10\,\text{fs}$. Its amplitude is $A_{0}$,
and $\omega_{\mathrm{pu}}$ is the central pump-pulse frequency, or
simply, the pump-pulse frequency ($\hbar\omega_{\mathrm{pu}}$ is
dubbed the pump-pulse photon energy).

We focus on the post-pump regime (time delay $t_{\mathrm{pr}}=40\,\text{fs}$)
for two fundamental reasons. First, from an experimental perspective,
measuring at large positive delays is easier because it requires less
effort to time-resolve the pump and probe envelopes, reducing the
need for ultra-precise synchronization. Second, and more importantly,
Floquet sidebands in the TR-ARPES spectrum require a time-periodic
driving term in $\Xi(\mathbf{k},t)$ (see Eq.~\ref{eq:S-full-t}).
After the pump pulse has subsided, the pump contribution to the self-energy
vanishes ($\Xi^{\mathrm{pu}}(\mathbf{k},t)=0$) and the time dependence
in the Hamiltonian cannot be attributed to the light field: it must
originate from the only other time-dependent term, namely the HF self-energy,
$\Xi^{\mathrm{HF}}(\mathbf{k},t)$, which indicates an oscillating
excitonic order parameter that persists after the pump is gone. Consequently,
the Floquet sidebands observed in the post-pump TR-ARPES spectrum
are uniquely due to the excitonic field, providing a clean signature
of coherent excitonic dynamics \citep{pareek2026driving}.

\begin{figure}
\centering{}\includegraphics[width=8cm]{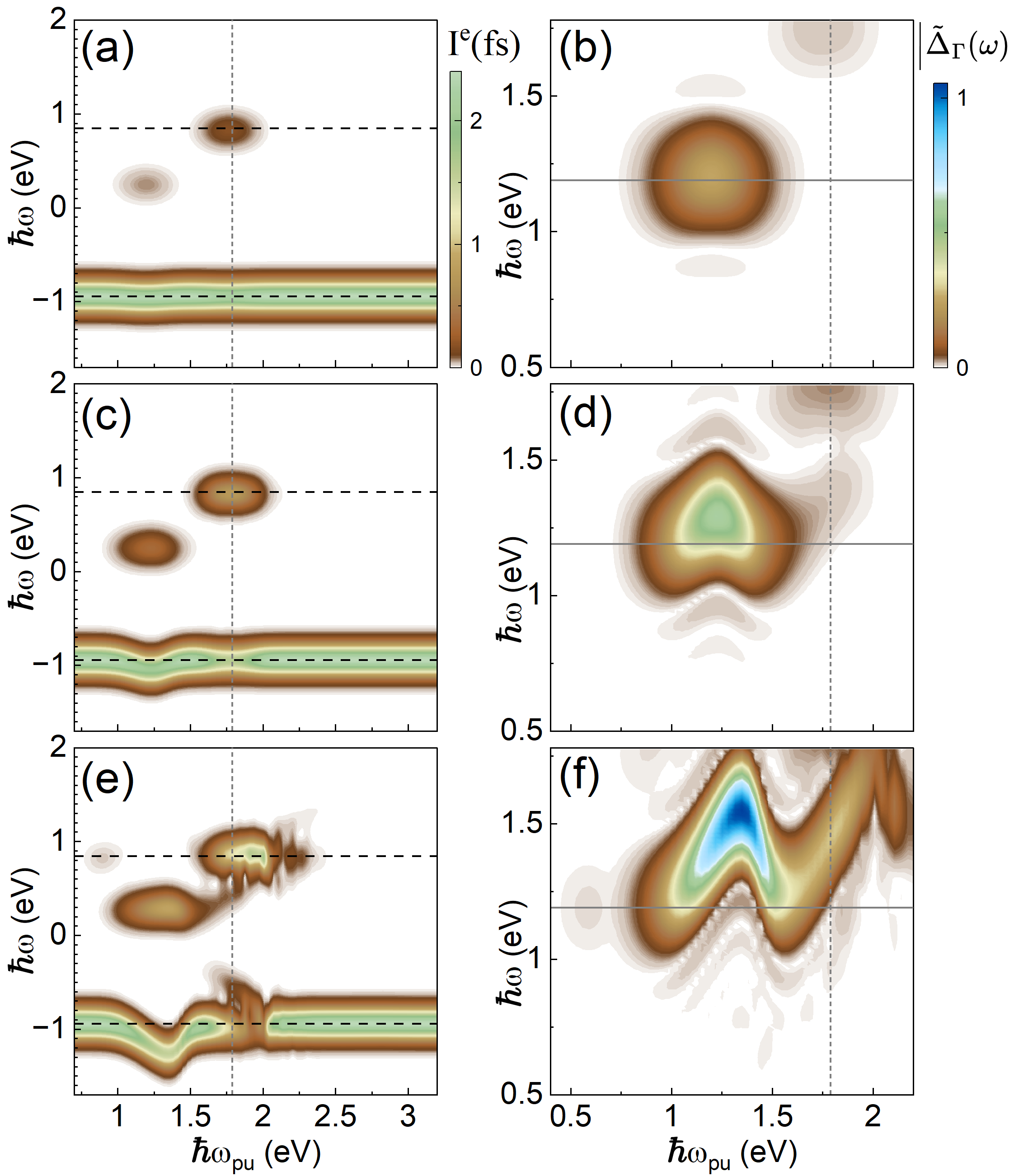}
\caption{Results for $V=6$ eV, semiconducting phase. (a), (c), (e) Post-pump
TR-ARPES signal at the $\Gamma$ point as a function of pump frequency
for pump amplitudes $A_{0}=0.1$, $0.25$, and $0.5\,\text{V}\cdot\text{fs}/\text{nm}$,
respectively. Horizontal dashed lines mark CB and VB energies at $\Gamma$;
vertical short-dashed lines indicate the pump frequency resonant with
the equilibrium band gap at $\Gamma$. (b), (d), (f) Corresponding
amplitude $|\tilde{\Delta}_{\Gamma}(\omega)|$ as a function of $\omega_{\mathrm{pu}}$,
and $\omega$. Horizontal gray lines indicate the equilibrium excitonic
energy from Fig.~\ref{fig:system_eq_pd_wexc}(c), and vertical short-dashed
lines are similar to the other panels. \label{fig:ARPES_G_semph}}
\end{figure}

Figure~\ref{fig:ARPES_G_semph} presents the post-pump TR-ARPES signal
for the semiconducting phase, with parameters $V=6\,\text{eV}$ and
$\kappa=0.2$. Panels (a), (c), and (e) show the post-pump TR-ARPES
signal at the $\Gamma$ point ($\mathbf{k}=\mathbf{0}$) as a function
of the pump frequency $\hbar\omega_{\mathrm{pu}}$ for different pump
intensities, identified by the amplitudes $A_{0}=0.1$, $0.25$, and
$0.5\,\text{V}\cdot\text{fs}/\text{nm}$, respectively. In these panels,
the horizontal dashed lines indicate the CB and VB energies at $\Gamma$.
Panels (b), (d), and (f) show the amplitude of the Fourier transform
of the normalized off-diagonal element of the Hartree--Fock self-energy,
$|\tilde{\Delta}_{\Gamma}(\omega)|$, as a function of $\omega_{\mathrm{pu}}$
and $\omega$, computed with the same pump-pulse parameters as their
companion panels (panels in the same row correspond to identical pump-pulse
parameters). The range of pump-pulse frequencies shown in these panels
is narrower than in the TR-ARPES panels because we are primarily interested
in the excitonic resonance frequencies that lie inside the band gap.
In these panels, the horizontal gray line indicates the equilibrium
excitonic energy, computed in Fig.~\ref{fig:system_eq_pd_wexc}(c).
In all panels of this figure, the vertical short-dashed lines indicate
the pump-pulse frequency that is in resonance with the equilibrium
band gap at $\Gamma$: $\hbar\omega_{\mathrm{pu}}=\varepsilon_{c}(\Gamma)-\varepsilon_{v}(\Gamma)$.
The time interval for all of the post-pump Fourier transformations
of this manuscript is from 25 to 55 fs, and the Hann windowing is
used to reduce the noise.

We now analyze each pair of panels (row) in detail.

\paragraph*{Panels (a) and (b): Weak pump intensity ($A_{0}=0.1\,\text{V}\cdot\text{fs}/\text{nm}$).}

In Fig.~\ref{fig:ARPES_G_semph}(a), a small exciton-field-induced
Floquet sideband appears, reaching its maximum intensity for $\hbar\omega_{\mathrm{pu}}=1.20\,\text{eV}$.
This value is very close to $1.19\,\text{eV}$, the corresponding
(same $V$ and same $\kappa$) equilibrium excitonic frequency obtained
from Fig.~\ref{fig:system_eq_pd_wexc}(c). For pump-pulse frequencies
near resonance with the equilibrium band gap, we observe residual
electron populations in the CB, that is, some electrons are excited
across the gap and remain there after the pump pulse has subsided.
For pump-pulse frequencies higher than the band-gap resonance, no
noticeable feature at the $\Gamma$ point emerges. Turning to Fig.~\ref{fig:ARPES_G_semph}(b),
$|\tilde{\Delta}_{\Gamma}(\omega)|$ exhibits its maximum at frequencies
$\omega$ close to the equilibrium excitonic frequency, and the corresponding
pump-pulse frequency is also almost in resonance with it, as indicated
in the previous panel. This demonstrates that at this weak pump intensity,
the excitonic frequency is not noticeably shifted from its equilibrium
value. Lowering the pump intensity further causes the post-pump excitonic
frequency and the equilibrium one to match (not shown).

\paragraph*{Panels (c) and (d): Intermediate pump intensity ($A_{0}=0.25\,\text{V}\cdot\text{fs}/\text{nm}$).}

In Fig.~\ref{fig:ARPES_G_semph}(c), we see that the pump-pulse frequency
that yields the strongest exciton-field-induced Floquet sideband,
has increased to $\hbar\omega_{\mathrm{pu}}=1.23\,\text{eV}$. Moreover,
at this excitonic resonance, the VB is shifted downward in energy,
a manifestation of the so-called Mexican-hat effect~\citep{parmenter1970superconductive,ma2021strongly,jia2022evidence,kogar2017signatures,gu2022dipolar,pareek2026driving}.
This effect arises from the coupling between the excitonic order parameter
and the single-particle band structure, which renormalizes the VB
dispersion near $\Gamma$. Importantly, the excitonic energy---defined
as the energy difference between the exciton-field-induced Floquet
sideband and the post-pump VB energy (the local maxima of the corresponding
TR-ARPES signal)---exceeds the pump-pulse photon energy $\hbar\omega_{\mathrm{pu}}$.
This observation is clarified in Fig.~\ref{fig:ARPES_G_semph}(d):
the maximum of $|\tilde{\Delta}_{\Gamma}(\omega)|$ occurs for $\hbar\omega_{\mathrm{pu}}=1.23\,\text{eV}$
but at $\hbar\omega=1.29\,\text{eV}$. Note that the latter value
is larger than $1.19\,\text{eV}$, the equilibrium excitonic frequency
computed in Fig.~\ref{fig:system_eq_pd_wexc}(c). This shift demonstrates
the effect of an intense pump pulse on the excitonic frequency. Because
the system is interacting and the Hamiltonian depends on the electronic
state through the HF self-energy, the pump-induced changes in the
electron distribution modify the effective attraction between electrons
and holes, thereby altering the excitonic resonance condition.

\paragraph*{Panels (e) and (f): Strong pump intensity ($A_{0}=0.5\,\text{V}\cdot\text{fs}/\text{nm}$).}

Increasing the intensity further magnifies all of the effects observed
in the previous panels. The pump-pulse frequency that gives the strongest
exciton-field-induced Floquet sideband increases further to $\hbar\omega_{\mathrm{pu}}=1.35\,\text{eV}$.
The Mexican-hat effect becomes noticeably stronger, as evidenced by
the more pronounced downward shift of the VB. Panel (f) shows that
the maximum of $|\tilde{\Delta}_{\Gamma}(\omega)|$ occurs for $\hbar\omega_{\mathrm{pu}}=1.35\,\text{eV}$
but at $\hbar\omega=1.55\,\text{eV}$, representing an even larger
shift than at intermediate intensity.

\begin{figure}
\centering{}\includegraphics[width=8cm]{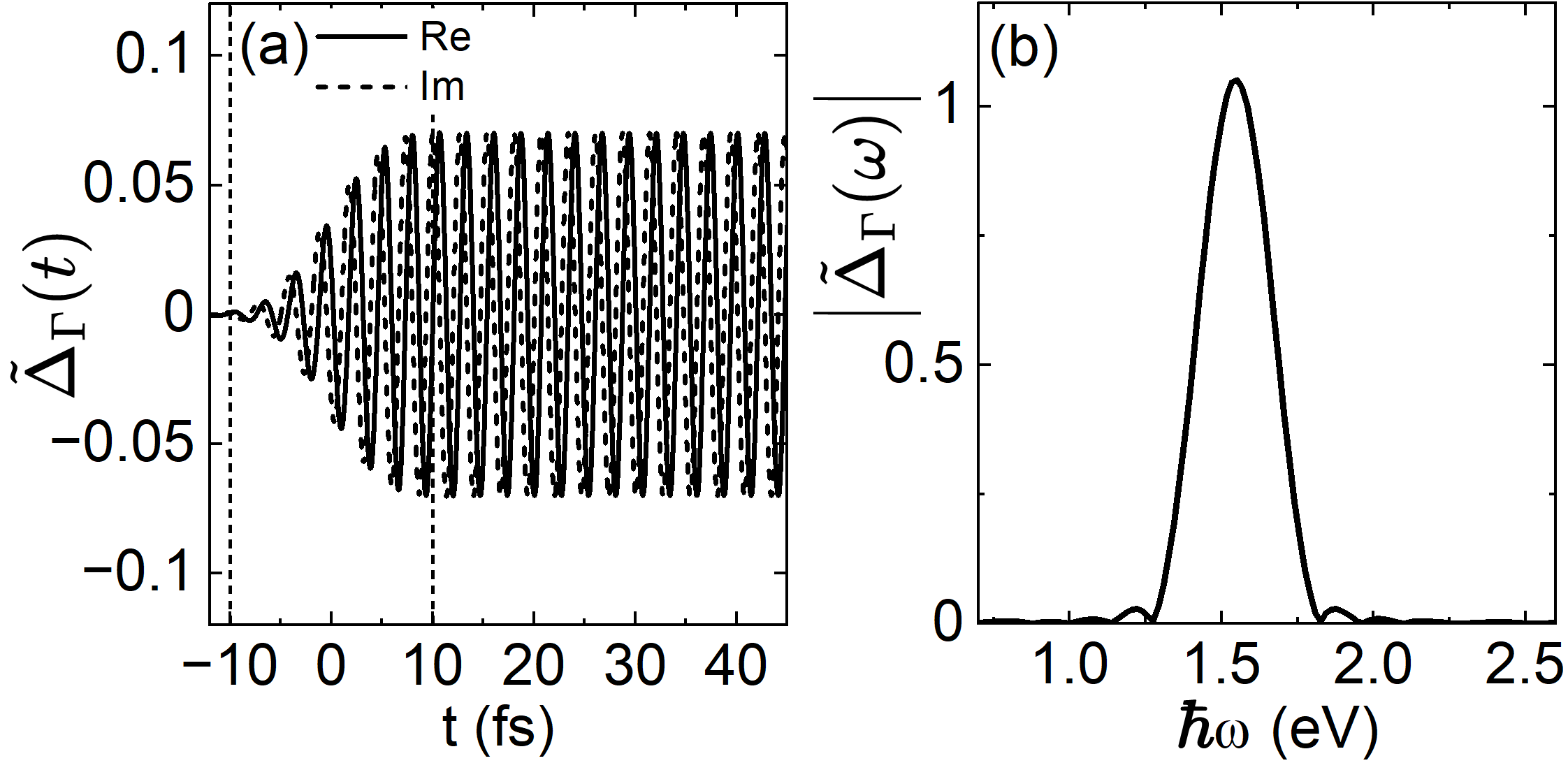} \caption{Results for $V=6$ eV, semiconducting phase. (a) Time evolution of
the real and imaginary parts of $\tilde{\Delta}_{\Gamma}(t)$ for
the high-intensity case ($A_{0}=0.5\,\text{V}\cdot\text{fs}/\text{nm}$)
with pump frequency $\hbar\omega_{\mathrm{pu}}=1.35\,\text{eV}$.
Vertical dashed lines indicate $t=\pm\tau_{\mathrm{pu}}=\pm10\,\text{fs}$,
marking the pump-pulse duration. (b) Corresponding post-pump amplitude
$|\tilde{\Delta}_{\Gamma}(\omega)|$ as a function of frequency, showing
the main exciton-field oscillation peak. \label{fig:Delta_G_t_semph}}
\end{figure}

To better clarify the behavior of the exciton-field oscillations,
we consider the high-intensity case ($A_{0}=0.5\,\text{V}\cdot\text{fs}/\text{nm}$)
and set the pump frequency to $\hbar\omega_{\mathrm{pu}}=1.35\,\text{eV}$,
which was identified in Fig.~\ref{fig:ARPES_G_semph}(e) as the value
yielding the strongest exciton-field-induced Floquet sideband. In
Fig.~\ref{fig:Delta_G_t_semph}(a), we plot $\tilde{\Delta}_{\Gamma}(t)$
as a function of time. The vertical dashed lines mark $t=\pm\tau_{\mathrm{pu}}=\pm10\,\text{fs}$,
indicating the temporal region where the pump pulse envelope is appreciably
nonzero, thereby clarifying when the pump pulse is turned on and off.
Both the real and imaginary parts of $\tilde{\Delta}_{\Gamma}(t)$
oscillate after the application of the pump pulse, while before its
application they are zero, as the system is initially in the semiconducting
phase with no preexisting excitonic order. To reveal the frequency
content of these post-pump oscillations, in Fig.~\ref{fig:Delta_G_t_semph}(b)
we plot the post-pump amplitude $|\tilde{\Delta}_{\Gamma}(\omega)|$
as a function of frequency. We clearly see the main peak, which determines
the dominant exciton-field oscillation frequency.

\begin{figure}
\centering{}\includegraphics[width=8cm]{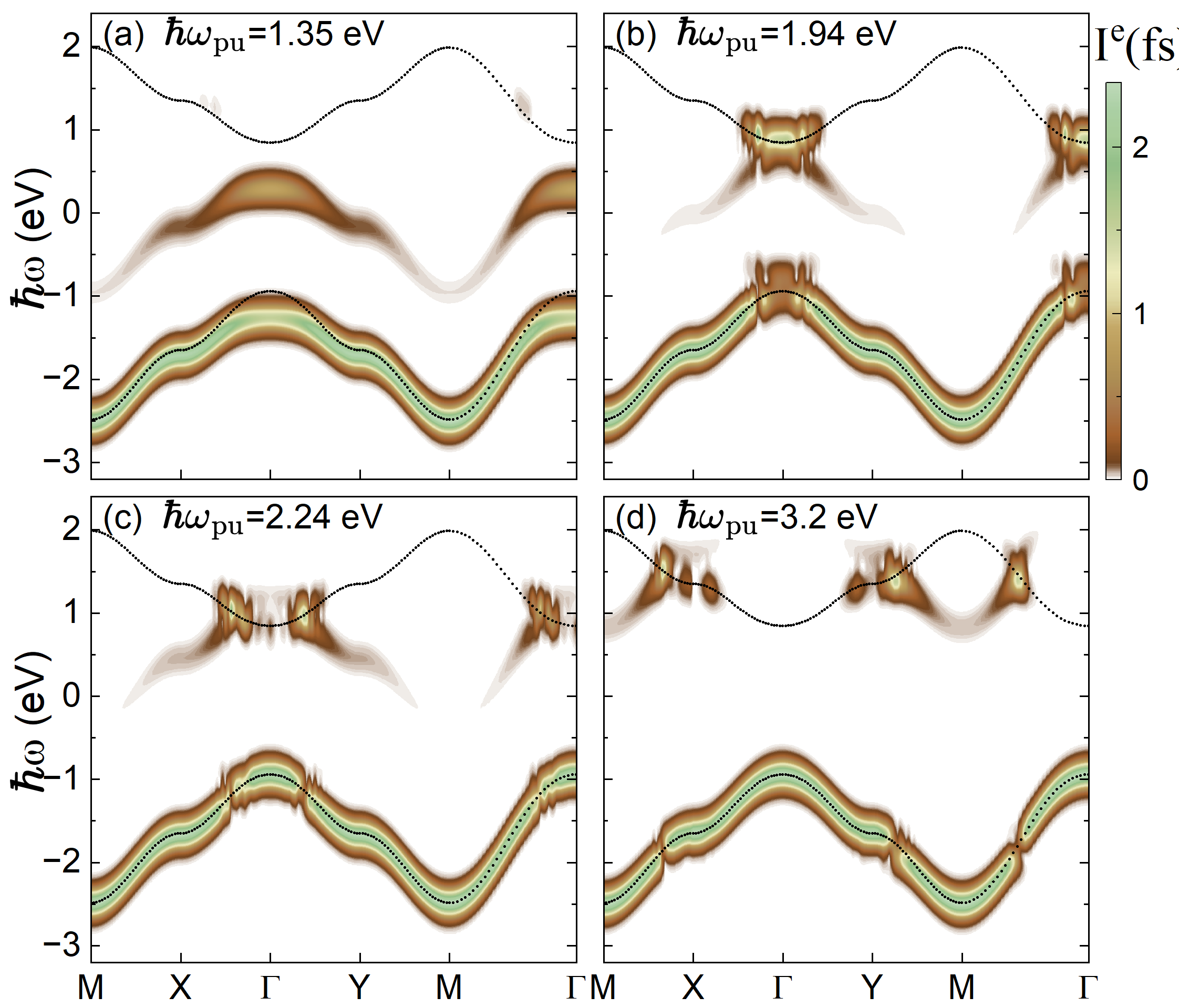} \caption{Results for $V=6$ eV, semiconducting phase. $\mathbf{k}$-resolved
post-pump TR-ARPES signal along the high-symmetry path in the BZ for
the large pump amplitude $A_{0}=0.5\,\text{V}\cdot\text{fs}/\text{nm}$,
corresponding to vertical cuts in Fig.~\ref{fig:ARPES_G_semph}(e).
(a) $\hbar\omega_{\mathrm{pu}}=1.35\,\text{eV}$, resonant with the
excitonic frequency, showing a strong exciton-field-induced Floquet
sideband parallel to the VB. (b) $\hbar\omega_{\mathrm{pu}}=1.94\,\text{eV}$,
slightly above the band gap at $\Gamma$, showing band-resonance-induced
sidebands and level splitting near $\Gamma$. (c) $\hbar\omega_{\mathrm{pu}}=2.24\,\text{eV}$,
with resonant $\mathbf{k}$-points farther away from $\Gamma$, yielding
weaker band-resonance-induced sidebands. (d) $\hbar\omega_{\mathrm{pu}}=3.2\,\text{eV}$,
with resonant $\mathbf{k}$-points very far from $\Gamma$, producing
very weak band-resonance-induced sidebands. The equilibrium band energies
are shown by dotted lines.\label{fig:arpes_k_semph}}
\end{figure}

In Fig.~\ref{fig:arpes_k_semph}, we plot the $\mathbf{k}$-resolved
post-pump TR-ARPES signal along the high-symmetry path in the BZ for
selected pump-pulse frequencies, all computed with the large pump
amplitude $A_{0}=0.5\,\text{V}\cdot\text{fs}/\text{nm}$. These cuts
correspond to vertical slices through the color plots shown in Fig.~\ref{fig:ARPES_G_semph}(e),
allowing us to examine the momentum-resolved structure of the post-pump
spectral features.

\paragraph*{Panel (a): $\hbar\omega_{\mathrm{pu}}=1.35\,\text{eV}$.}

This pump frequency was identified in Fig.~\ref{fig:ARPES_G_semph}(e)
as the one giving the strongest exciton-field-induced Floquet sideband.
Here, we observe a sideband induced by the excitonic field that appears
almost everywhere parallel to the VB, which is a characteristic signature
of a Floquet sideband. However, this sideband is stronger near $\Gamma$
and becomes significantly weaker away from the zone center. This momentum
dependence has two origins. First, the HF self-energy itself turns
out to be larger in magnitude at $\Gamma$ than at other $\mathbf{k}$-points,
leading to a stronger effective driving. Second, and more importantly,
the oscillation frequency of the HF self-energy at $\Gamma$ is close
to the local equilibrium band gap energy at that $\mathbf{k}$-point;
this small-detuning condition enhances the sideband intensity. The
latter factor turns out to be much more important; we will clarify
this point further when we analyze the local Coulomb interaction case.
We note that there are residual electronic excitations at $\mathbf{k}$-points
that satisfy a two-photon resonance condition with the pump pulse,
but their contribution is too small to be visible on the scale of
this figure.

\paragraph*{Panel (b): $\hbar\omega_{\mathrm{pu}}=1.94\,\text{eV}$.}

This pump frequency is slightly above the direct band gap resonance
at $\Gamma$. In this case, we observe residual excitations at the
resonant $\mathbf{k}$-points, which give rise to some Floquet sidebands
that have a different origin from those of the previous panel. For
pump-pulse photon energies larger than the band gap, and with an intense
enough pump pulse, we observe a Floquet sideband located approximately
at an energy corresponding to $\hbar\omega_{\mathrm{pu}}$ above the
main band energies. Whenever the pump-photon energy exceeds the energy
gap, there exist some $\mathbf{k}$-points in the BZ whose band gap
is in resonance with the pump frequency. Consequently, a residual
excitation remains in the system after the pump pulse has subsided.
These residual excitations possess coherence terms that oscillate
with their own energy gap, which is equal to $\hbar\omega_{\mathrm{pu}}$.
These coherences contribute to the HF self-energy and provide the
oscillatory Hamiltonian term which is necessary for the emergence
of Floquet sidebands. As such, these sidebands are not strictly induced
by excitons; we refer to them as band-resonance-induced sidebands
to distinguish them from the genuine exciton-field-induced Floquet
features.

It should be noted that this pump-photon energy also excites some
excitonic coherences, and therefore the induced sidebands have a mixed
origin, partly from the excitonic field and partly from the band-gap
resonance. This hybrid nature can be verified by a detailed analysis
of $\tilde{\Delta}_{\Gamma}(\omega)$ for this case (not shown). Finally,
the resonant driving condition causes a level splitting near $\Gamma$,
an effect that was also visible in Fig.~\ref{fig:ARPES_G_semph}(e).

\paragraph*{Panel (c): $\hbar\omega_{\mathrm{pu}}=2.24\,\text{eV}$.}

Here, the resonant $\mathbf{k}$-points are located farther away from
$\Gamma$ than in the previous case. The band-resonance-induced Floquet
sidebands emerge clearly, and as explained earlier, they appear at
an energy approximately $\hbar\omega_{\mathrm{pu}}$ above the VB.
These sidebands are weaker than those in the preceding panels.

\paragraph*{Panel (d): $\hbar\omega_{\mathrm{pu}}=3.2\,\text{eV}$.}

At this high pump frequency, the resonant $\mathbf{k}$-points are
very far from $\Gamma$, and the corresponding band-resonance-induced
Floquet sidebands are very weak. When the pump is tuned far away from
the excitonic resonance, the induced sidebands become progressively
weaker and eventually faint away.

\begin{figure}
\centering{}\includegraphics[width=8cm]{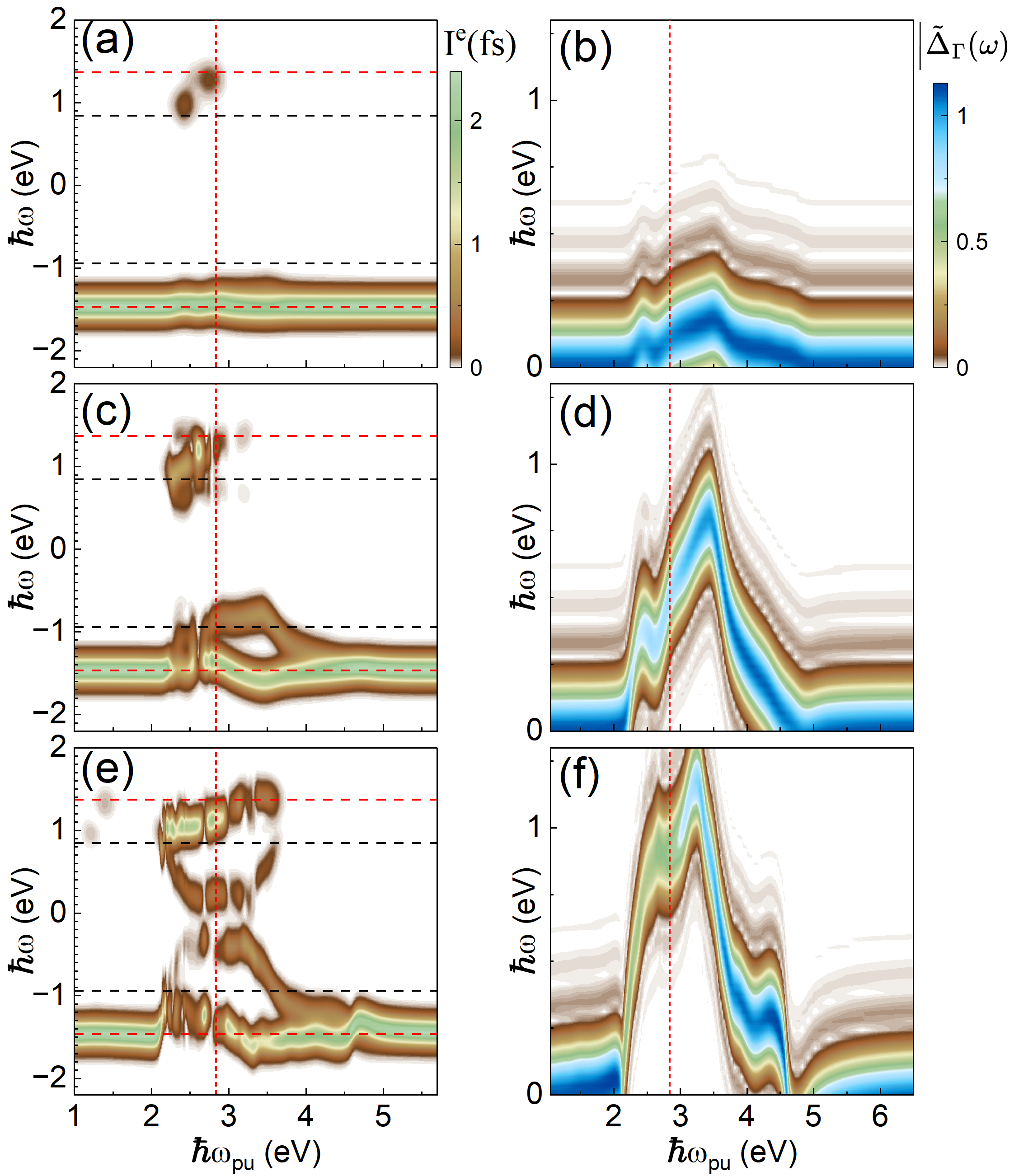}
\caption{Results for $V=15$ eV, excitonic-insulator phase. (a), (c), (e) Post-pump
TR-ARPES signal at the $\Gamma$ point as a function of pump frequency
$\hbar\omega_{\mathrm{pu}}$ for the excitonic-insulator phase ($V=6\,\text{eV}$,
$\kappa=0.2$) and pump amplitudes $A_{0}=0.1$, $0.25$, and $0.5\,\text{V}\cdot\text{fs}/\text{nm}$,
respectively. Black horizontal dashed lines mark the noninteracting
CB and VB energies at $\Gamma$; red horizontal dashed lines mark
the \emph{interacting} band energies. (b), (d), (f) Corresponding
amplitude $|\tilde{\Delta}_{\Gamma}(\omega)|$ as a function of $\omega_{\mathrm{pu}}$,
and $\omega$. Vertical red dashed lines indicate the pump frequency
resonant with the equilibrium band gap of the interacting system.
\label{fig:ARPES_G_excph}}
\end{figure}

\subsection{Out-of-Equilibrium: Excitonic-Insulator Phase}

We now turn to the excitonic-insulator phase. We choose parameters
$V=15\,\text{eV}$ and $\kappa=0.2$, which place the system well
inside the excitonic-insulator phase as established in Fig.~\ref{fig:system_eq_pd_wexc}(b).

Figure~\ref{fig:ARPES_G_excph} is the counterpart of Fig.~\ref{fig:ARPES_G_semph}
for the excitonic-insulator phase. As in the semiconducting case,
panels (a), (c), and (e) show the post-pump TR-ARPES signal at the
$\Gamma$ point as a function of the pump frequency $\hbar\omega_{\mathrm{pu}}$
for three different pump intensities $A_{0}=0.1$, $0.25$, and $0.5\,\text{V}\cdot\text{fs}/\text{nm}$,
respectively. Panels (b), (d), and (f) show the corresponding amplitude
$|\tilde{\Delta}_{\Gamma}(\omega)|$ as a function of $\omega_{\mathrm{pu}}$,
and $\omega$, computed with the same pump-pulse parameters as their
companion panels. In panels (a), (c), and (e), the black horizontal
dashed lines indicate the CB and VB energies of the noninteracting
system, while the red horizontal dashed lines indicate those of the
interacting system. Because of exciton formation in equilibrium, the
band energies must be computed by diagonalizing the full interacting
Hamiltonian, and these red lines mark the \emph{interacting} band
positions. The vertical red dashed line indicates the equilibrium
resonance condition using the \emph{interacting} bands.

We now analyze each pair of panels in detail.

\paragraph*{Panels (a) and (b): Weak pump intensity ($A_{0}=0.1\,\text{V}\cdot\text{fs}/\text{nm}$).}

Well before reaching the resonance condition, no noticeable features
appear in the TR-ARPES signal. As the pump frequency approaches the
resonance, we observe some residual electrons appearing in the CB.
This indicates that the pump pulse has dissociated some of the equilibrium
excitons, causing the bands to shift slightly and move toward the
noninteracting band positions. Notably, almost no exciton-field-induced
Floquet sideband is visible. Panel (b) confirms this observation:
there is only a very low-frequency component in $|\tilde{\Delta}_{\Gamma}(\omega)|$
for pump-pulse frequencies near and above the resonance at $\Gamma$,
indicating that no noticeable coherent high-frequency excitonic oscillation
persists after the pump.

\paragraph*{Panels (c) and (d): Intermediate pump intensity ($A_{0}=0.25\,\text{V}\cdot\text{fs}/\text{nm}$).}

Again, away from the resonance, no features are visible. Near the
resonance, we obtain residual electrons in the CB, with a larger population
than in the weak-intensity case. An exciton-field-induced sideband
appears, running parallel to both the VB and the occupied portions
of the CB. Panel (d) confirms that for pump-pulse photon energies
near and above the resonance, there exists a well-defined excitonic
field frequency, manifested as a local maximum in $|\tilde{\Delta}_{\Gamma}(\omega)|$.
Note that for very large pump photon energies exceeding the overall
bandwidth, no resonance occurs, and consequently no features are observed.

\paragraph*{Panels (e) and (f): Strong pump intensity ($A_{0}=0.5\,\text{V}\cdot\text{fs}/\text{nm}$).}

The behavior is qualitatively similar to the intermediate-intensity
case, but the effects are generally stronger. The bending of the bands
toward the noninteracting positions near the resonance is more pronounced.
Additionally, a residual signal appears when the pump-pulse photon
energy is approximately half the band gap, arising from a two-photon
resonance process. These panels clearly demonstrate how strongly the
exciton-field-induced sideband depends on the pump-pulse intensity.

\begin{figure}
\centering{}\includegraphics[width=8cm]{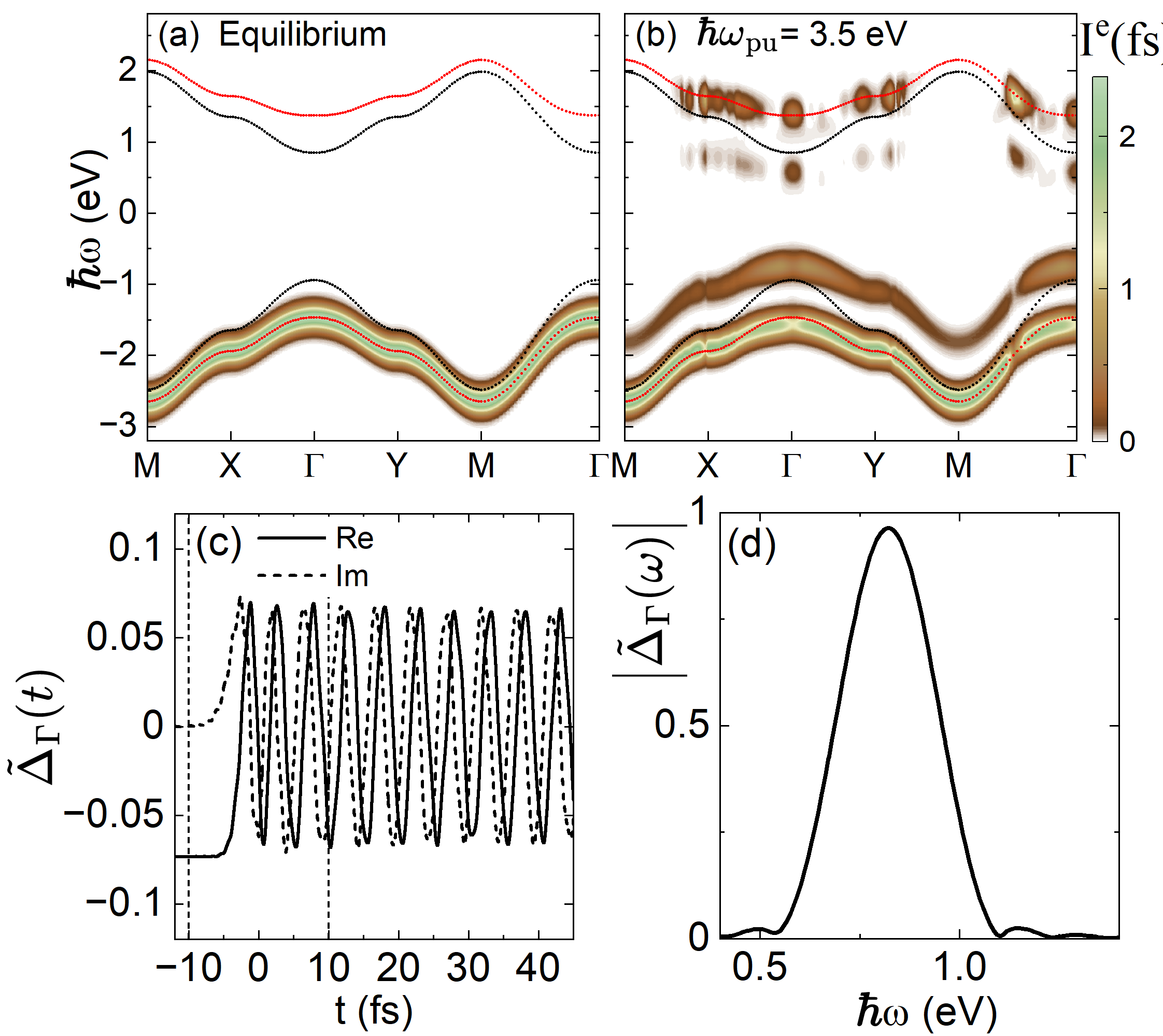} \caption{Results for $V=15$ eV, excitonic-insulator phase. (a) Equilibrium
$\mathbf{k}$-resolved ARPES signal along the high-symmetry path in
the BZ for the excitonic-insulator phase (no pump). (b) Post-pump
$\mathbf{k}$-resolved TR-ARPES signal for $\hbar\omega_{\mathrm{pu}}=3.5\,\text{eV}$
and $A_{0}=0.5\,\text{V}\cdot\text{fs}/\text{nm}$, showing exciton-field-induced
sidebands parallel to the occupied CB and VB. In panels (a) and (b),
red dotted lines indicate the interacting band structure from the
self-consistent procedure (Appendix~\ref{sec:SC_EQ}); black dotted
lines indicate the noninteracting bands. (c) Time evolution of $\tilde{\Delta}_{\Gamma}(t)$
for the same parameters as panel (b), showing reduction and oscillations
upon pumping. (d) Post-pump Fourier transform $|\tilde{\Delta}_{\Gamma}(\omega)|$,
revealing the main excitonic oscillation frequency. \label{fig:arpes_k_excph}}
\end{figure}

Figure~\ref{fig:arpes_k_excph} presents momentum-resolved results
for the excitonic-insulator phase.

In Fig.~\ref{fig:arpes_k_excph}(a), we show the equilibrium $\mathbf{k}$-resolved
ARPES signal along the high-symmetry path in the BZ (no pump). The
signal is centered around the lower \emph{interacting} energy band.
The \emph{interacting} bands are indicated by the red dotted lines,
and represent the eigenvalues of the full Hamiltonian after the self-consistent
procedure described in the Appendix~\ref{sec:SC_EQ}. For comparison,
the noninteracting bands are indicated by the black dotted lines.
The renormalization of the band structure due to exciton formation
is clearly visible.

In Fig.~\ref{fig:arpes_k_excph}(b), we show the $\mathbf{k}$-resolved
post-pump TR-ARPES signal along the high-symmetry path for pump frequency
$\hbar\omega_{\mathrm{pu}}=3.5\,\text{eV}$ and pump amplitude $A_{0}=0.5\,\text{V}\cdot\text{fs}/\text{nm}$.
There are $\mathbf{k}$-points that are in resonance with this pump-pulse
frequency, and these points host post-pump electrons in their CB.
The broken excitonic bonds induce coherent oscillations, giving rise
to exciton-field-induced sidebands that run parallel to both the occupied
CB and the VB. Additionally, we observe occupied CB states at $\mathbf{k}$-points
that are not in direct resonance with the pump frequency. This occurs
because during the application of the pump pulse, the HF self-energy
is finite and oscillating, which can generate additional resonances
involving the sum or difference of these oscillation frequencies with
the pump frequency.

In Fig.~\ref{fig:arpes_k_excph}(c) and (d), we examine the excitonic
order parameter dynamics. Panel (c) shows $\tilde{\Delta}_{\Gamma}(t)$
as a function of time for the same parameters. Initially, in equilibrium,
$\tilde{\Delta}_{\Gamma}$ is finite, reflecting the presence of the
excitonic order parameter in the excitonic-insulator phase. Upon application
of the pump pulse, its absolute value is reduced, and both the real
and imaginary parts exhibit oscillations, indicating that the pump
has partially suppressed but not completely destroyed the excitonic
order, leaving a coherently oscillating component. Panel (d) displays
the amplitude of the post-pump Fourier transform, $|\tilde{\Delta}_{\Gamma}(\omega)|$,
which reveals a main peak at a characteristic excitonic frequency.
This confirms that coherent excitonic oscillations persist after the
pump pulse in the excitonic-insulator phase, with a reduced amplitude
compared to the equilibrium value of $\tilde{\Delta}_{\Gamma}$.

\begin{figure}
\centering{}\includegraphics[width=8cm]{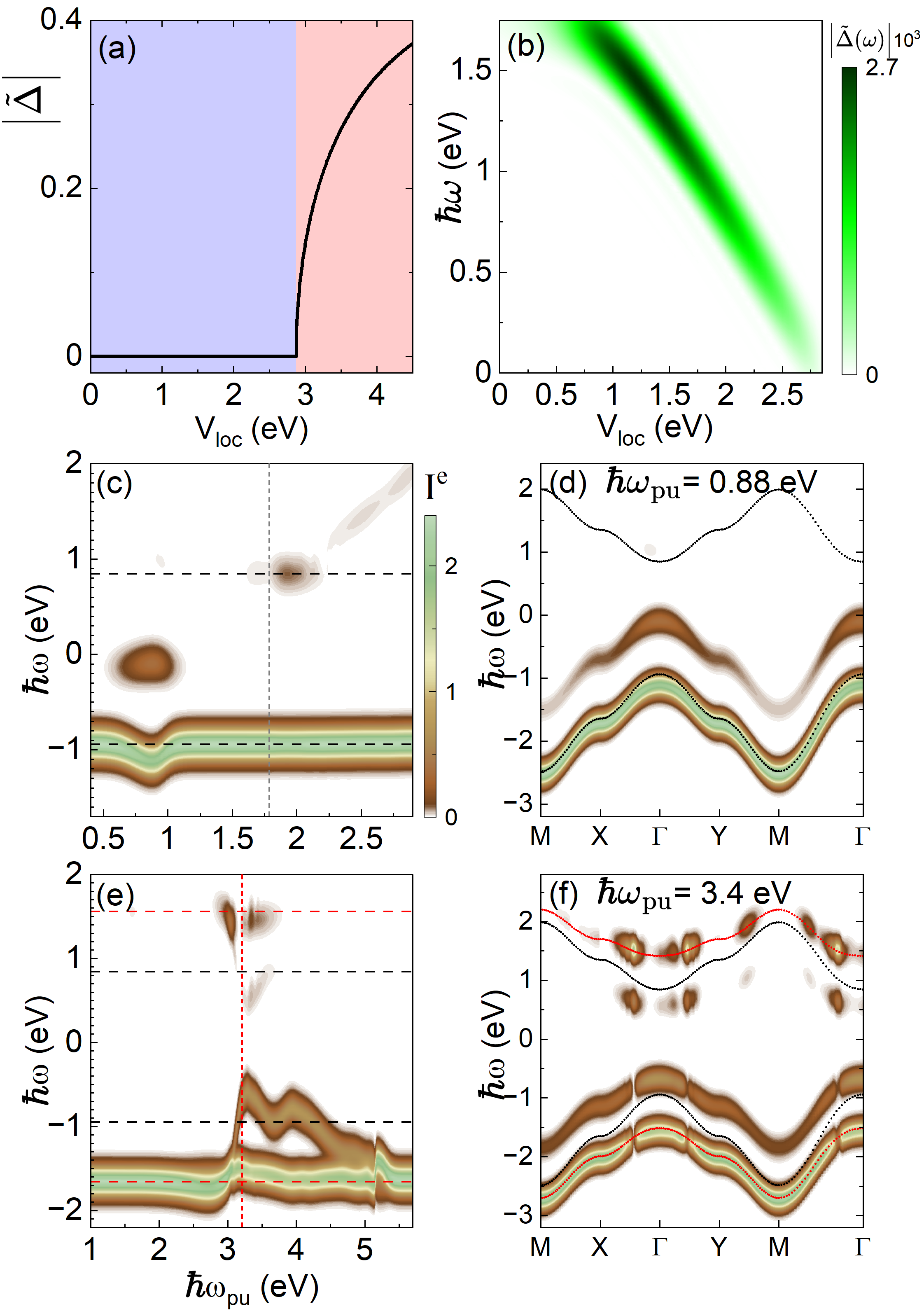} \caption{Results for the local interaction. (a) Equilibrium $\tilde{\Delta}$
as a function of the local Coulomb interaction strength $V_{\mathrm{loc}}$,
showing the transition between semiconducting and excitonic-insulator
phases at $V_{\mathrm{loc}}\approx2.87\,\text{eV}$. (b) $|\tilde{\Delta}(\omega)|$
as a function of $V_{\mathrm{loc}}$ and frequency, giving the equilibrium
excitonic frequency in the local interaction limit, for the same protocol
as in Figs.~\ref{fig:system_eq_pd_wexc}(c) and (d). (c) Post-pump
TR-ARPES signal at $\Gamma$ for the semiconducting phase ($V_{\mathrm{loc}}=2\,\text{eV}$)
with $A_{0}=0.5\,\text{V}\cdot\text{fs}/\text{nm}$, counterpart of
Fig.~\ref{fig:ARPES_G_semph}(e). (d) $\mathbf{k}$-resolved post-pump
TR-ARPES signal for $\hbar\omega_{\mathrm{pu}}=0.88\,\text{eV}$ and
$A_{0}=0.5\,\text{V}\cdot\text{fs}/\text{nm}$ in the semiconducting
phase, counterpart of Fig.~\ref{fig:arpes_k_semph}(a). (e) Post-pump
TR-ARPES signal at $\Gamma$ for the excitonic-insulator phase ($V_{\mathrm{loc}}=4\,\text{eV}$)
with $A_{0}=0.5\,\text{V}\cdot\text{fs}/\text{nm}$, counterpart of
Fig.~\ref{fig:ARPES_G_excph}(e). (f) $\mathbf{k}$-resolved post-pump
TR-ARPES signal for $\hbar\omega_{\mathrm{pu}}=3.4\,\text{eV}$ and
$A_{0}=0.5\,\text{V}\cdot\text{fs}/\text{nm}$ in the excitonic-insulator
phase, counterpart of Fig.~\ref{fig:arpes_k_excph}(b).\label{fig:v_loc}}
\end{figure}

\subsection{Local Coulomb Interaction Limit}

In this section, we analyze the limit of local interaction. This limit
is obtained by setting both $V\rightarrow\infty$ and $\kappa\rightarrow\infty$
in Eq.~\ref{eq:V_k_LR}, such that $V(\mathbf{k})\rightarrow V_{\mathrm{loc}}=V/\kappa^{2}$
remains finite and becomes momentum independent. In this limit, the
HF self-energy becomes $\mathbf{k}$-independent, which simplifies
the numerical calculations.

In Fig.~\ref{fig:v_loc}(a), we study the equilibrium phase diagram
by plotting the equilibrium value of $\tilde{\Delta}_{\Gamma}=\tilde{\Delta}$
as a function of $V_{\mathrm{loc}}$. There is a critical local Coulomb
interaction strength of approximately $2.87\,\text{eV}$. Below this
critical value, the system is in the semiconducting phase; above it,
the system enters the excitonic-insulator phase. Applying the same
protocol as in Figs.~\ref{fig:system_eq_pd_wexc}(c) and (d), we
plot $|\tilde{\Delta}(\omega)|$ as a function of $V_{\mathrm{loc}}$
and frequency in Fig.~\ref{fig:v_loc}(b), thereby obtaining the
equilibrium excitonic frequency in this local interaction regime.

In Figs.~\ref{fig:v_loc}(c) and (d), we consider $V_{\mathrm{loc}}=2\,\text{eV}$,
which corresponds to the semiconducting phase. Panel (c) shows the
post-pump TR-ARPES signal at the $\Gamma$ point ($\mathbf{k}=\mathbf{0}$)
as a function of the pump frequency $\hbar\omega_{\mathrm{pu}}$ for
the amplitude $A_{0}=0.5\,\text{V}\cdot\text{fs}/\text{nm}$. This
figure is the counterpart of Fig.~\ref{fig:ARPES_G_semph}(e). The
qualitative behavior is similar, but quantitatively it reflects the
$\mathbf{k}$-independent nature of the interaction. Panel (d) displays
the $\mathbf{k}$-resolved post-pump TR-ARPES signal along the high-symmetry
path for pump frequency $\hbar\omega_{\mathrm{pu}}=0.88\,\text{eV}$
and pump amplitude $A_{0}=0.5\,\text{V}\cdot\text{fs}/\text{nm}$.
This figure is the counterpart of Fig.~\ref{fig:arpes_k_semph}(a).
Again, the qualitative behavior is similar, while the quantitative
details differ. Also in this case, the excitonic-field-induced Floquet
sideband is stronger near the $\Gamma$-point than at other momenta.
This behavior clarifies what we have already stated for the $\mathbf{k}$-dependent
interaction: it is the proximity of the excitonic frequency to the
energy band gap at $\Gamma$ (the small-detuning condition) that causes
a stronger sideband close to $\Gamma$ and not the $\mathbf{k}$-dependence
of the interaction.

In Figs.~\ref{fig:v_loc}(e) and (f), we consider $V_{\mathrm{loc}}=4\,\text{eV}$,
which corresponds to the excitonic-insulator phase. Panel (e) shows
the post-pump TR-ARPES signal at the $\Gamma$ point ($\boldsymbol{k}=0$)
as a function of the pump frequency $\hbar\omega_{\mathrm{pu}}$ for
the amplitude $A_{0}=0.5\,\text{V}\cdot\text{fs}/\text{nm}$. This
figure is the counterpart of Fig.~\ref{fig:ARPES_G_excph}(e). Panel
(f) shows the $\mathbf{k}$-resolved post-pump TR-ARPES signal along
the high-symmetry path for pump frequency $\hbar\omega_{\mathrm{pu}}=3.4\,\text{eV}$
and pump amplitude $A_{0}=0.5\,\text{V}\cdot\text{fs}/\text{nm}$.
This figure is the counterpart of Fig.~\ref{fig:arpes_k_excph}(b).
Once more, the qualitative behavior is similar, but quantitatively
it is different.

\section{Summary and Conclusions\label{sec:Conclusions}}

In this work, we have developed a theoretical framework based on the
Dynamical Projective Operatorial Approach (DPOA) to investigate the
time-resolved ARPES response of pumped excitonic systems. By incorporating
Coulomb electron--electron interactions at the Hartree-Fock level,
our formalism captures the formation of excitonic bound states both
in equilibrium and under the influence of an external pump pulse.
We have applied this framework to a prototypical two-dimensional two-band
semiconductor model, systematically exploring the equilibrium phase
diagram and the out-of-equilibrium dynamics for both the semiconducting
and the excitonic-insulator phases.

Our results demonstrate that pump-induced coherent oscillations of
the excitonic order parameter persist after the pump pulse subsides
and give rise to clear excitonic-field-driven Floquet sidebands in
the TR-ARPES signal. These Floquet sidebands are distinct from those
originating from the pump laser field itself and provide a direct
fingerprint of excitonic coherence.

For the semiconducting phase, where no excitons exist in equilibrium,
we showed that resonant pumping at the excitonic frequency generates
coherent excitonic oscillations that manifest as Floquet sidebands
parallel to the VB in the post-pump TR-ARPES spectrum. The Mexican-hat
effect emerges for sufficiently strong pump intensities, and we distinguished
genuine exciton-field-induced sidebands from band-resonance-induced
sidebands, which arise from residual coherences at momentum points
resonant with the pump frequency.

For the excitonic-insulator phase, where excitonic order is already
present in equilibrium, we found that the pump pulse partially melts
the excitonic order, shifting the bands toward their \emph{noninteracting}
positions while simultaneously generating exciton-field-induced sidebands.
The time evolution of the excitonic order parameter reveals that the
pump reduces its magnitude but leaves a coherently oscillating component,
which is responsible for the observed Floquet features. Our analysis
also showed that these effects become more pronounced with increasing
pump intensity.

By studying the local Coulomb interaction limit, we isolated the role
of momentum-dependent interactions. Qualitatively the same physical
phenomena are observed also in this case. In particular, we found
that even though the HF self-energy becomes $\mathbf{k}$-independent,
the exciton-field-induced sidebands remain strongest at $\Gamma$,
confirming that the primary mechanism for this momentum dependence
is the small detuning between the excitonic frequency and the local
band gap, rather than the momentum structure of the HF self-energy
itself.

Our findings corroborate the recent experimental observations of Ref.~\citep{pareek2026driving},
where exciton-driven Floquet-like sidebands were reported in monolayer
transition metal dichalcogenides. The agreement between our theoretical
predictions and those experimental results validates both the physical
picture of exciton-field-induced sidebands and the efficacy of the
DPOA framework in capturing the essential dynamics of pumped excitonic
systems.

Our work opens several avenues for future research. The DPOA framework
can be extended to more complex models with realistic band structures.
Furthermore, the connection between pump-induced excitonic oscillations
and Floquet engineering suggests new possibilities for controlling
material properties on ultrafast timescales, potentially leading to
the realization of novel non-equilibrium phases of matter. By establishing
a robust theoretical description of excitonic-field-driven Floquet
sidebands in TR-ARPES, our work provides a foundation for interpreting
ongoing and future experiments and for designing strategies to harness
excitonic coherence in next-generation optoelectronic and quantum
technologies.
\begin{acknowledgments}
The authors thank C. Giannetti for insightful discussions and acknowledge
support by MUR under Project PNRR MUR Missione 4 (SPOKE 2) TOPQIN
``TOPological Qubit In driveN and reconfigurable heterostructures''.
\end{acknowledgments}

\appendix

\section{Self-consistent calculations at equilibrium\label{sec:SC_EQ}}

\begin{figure}
\centering{}\includegraphics[width=1\columnwidth]{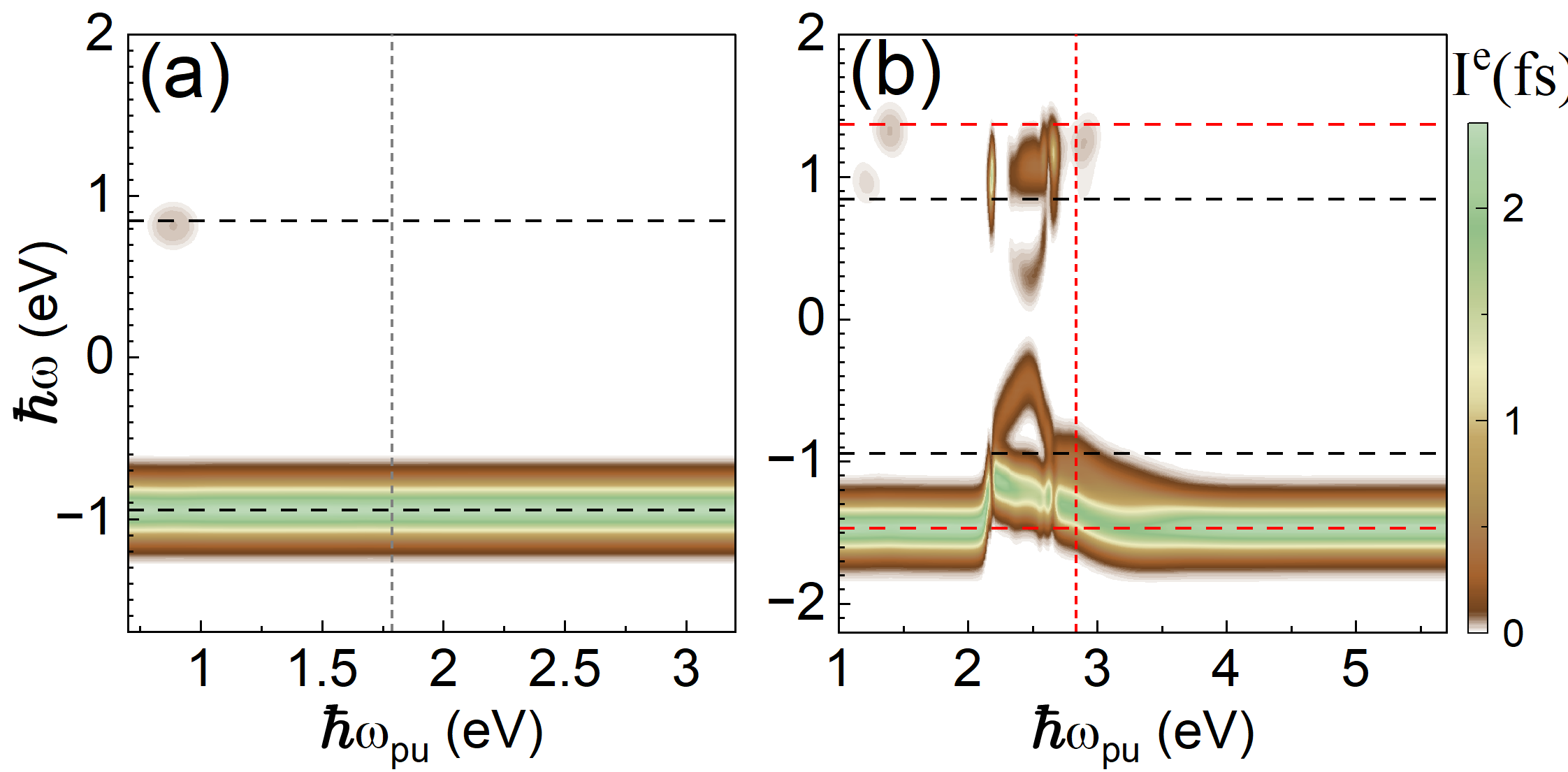}
\caption{Influence of the dipole strength $D$ on the post-pump TR-ARPES signal
at $\Gamma$. (a) Post-pump TR-ARPES signal at the $\Gamma$ point
for the semiconducting phase with $D=0$, counterpart of Fig.~\ref{fig:ARPES_G_semph}(e).
The excitonic resonance disappears. (b) Post-pump TR-ARPES signal
at the $\Gamma$ point for the excitonic-insulator phase with $D=0$,
counterpart of Fig.~\ref{fig:ARPES_G_excph}(e). The main excitonic
sideband persists but becomes much weaker and qualitatively different.\label{fig:Ds}}
\end{figure}

The full equilibrium Hamiltonian obtained after the HF decoupling
reads 
\begin{equation}
H=\sum_{\mathbf{k}}c^{\dagger}\left(\mathbf{k}\right)\cdot\Xi\left(\mathbf{k}\right)\cdot c\left(\mathbf{k}\right),
\end{equation}
with 
\begin{equation}
\Xi\left(\mathbf{k}\right)=\Xi^{0}\left(\mathbf{k}\right)+\Xi^{HF}\left(\mathbf{k}\right).
\end{equation}
Since $\Xi^{HF}$ itself depends on the single-particle density matrix,
$\rho$, a self-consistent procedure is required. We work in the band
basis, i.e., the basis where $\Xi^{0}\left(\mathbf{k}\right)$ is
diagonal. The iterative scheme proceeds as follows:
\begin{enumerate}
\item Start with an initial guess $\rho^{\left(0\right)}\left(\mathbf{k}\right)$
for all $\mathbf{k}$.
\item Construct $\Xi\left(\mathbf{k}\right)$ from the current $\rho\left(\mathbf{k}\right)$
using Eq.~\ref{eq:S-HF}.
\item Diagonalize $\Xi\left(\mathbf{k}\right)$ via a unitary transformation
$U_{\mathbf{k}}$ to obtain its eigenvalues, stored in a vector $\bar{\xi}\left(\mathbf{k}\right)$:
\begin{equation}
U^{\dagger}_{\mathbf{k}}\cdot\Xi\left(\mathbf{k}\right)\cdot U_{\mathbf{k}}=\text{diag}\left[\bar{\xi}\left(\mathbf{k}\right)\right].
\end{equation}
\item In this diagonal basis, the density matrix is diagonal with entries
given by the Fermi--Dirac distribution at inverse temperature $\beta=\frac{1}{k_{\mathrm{B}}T}$:
\begin{equation}
\bar{\rho}\left(\mathbf{k}\right)=\text{diag}\left[f_{\beta}\left(\bar{\xi}\left(\mathbf{k}\right)\right)\right].
\end{equation}
\item Transform back to the original band basis: 
\begin{equation}
\rho^{\left(1\right)}\left(\mathbf{k}\right)=U_{\mathbf{k}}\cdot\bar{\rho}\left(\mathbf{k}\right)\cdot U^{\dagger}_{\mathbf{k}}.
\end{equation}
\item If $\rho^{\left(1\right)}\left(\mathbf{k}\right)$ is sufficiently
close to $\rho^{\left(0\right)}\left(\mathbf{k}\right)$ for all $\mathbf{k}$,
the solution is found. Otherwise, mix the two according to $\rho^{\left(0\right)}\left(\mathbf{k}\right)\rightarrow r_{\text{mix}}\rho^{\left(0\right)}\left(\mathbf{k}\right)+\left(1-r_{\text{mix}}\right)\rho^{\left(1\right)}\left(\mathbf{k}\right)$
with a mixing parameter $0\le r_{\text{mix}}<1$, and return to step
2.
\end{enumerate}
The final converged density matrix is denoted $\rho^{\text{eq}}\left(\mathbf{k}\right)$.

\section{The effects of asymmetry induced by the local dipole moment\label{sec:The-effects-of-Ds}}

In this appendix, we examine how the value of the dipole parameter
$D$ affects the TR-ARPES spectra.

Fig.~\ref{fig:Ds}(a) is the counterpart of Fig.~\ref{fig:ARPES_G_semph}(e)
for $D=0$, corresponding to the semiconducting phase. Removing the
asymmetry by setting $D=0$ makes the excitonic resonance disappear
entirely. No exciton-field-induced Floquet sideband is visible in
the post-pump TR-ARPES signal, indicating that a symmetry-breaking
is essential for the generation and detection of coherent excitonic
oscillations in the semiconducting phase.

Fig.~\ref{fig:Ds}(b) is the counterpart of Fig.~\ref{fig:ARPES_G_excph}(e)
for $D=0$, corresponding to the excitonic-insulator phase. In contrast
to the semiconducting case, the main excitonic sideband persists even
when $D=0$. However, it becomes much weaker and exhibits different
features compared to the case with finite $D$. This indicates that
while excitonic order exists in equilibrium in the excitonic-insulator
phase, the dipole moment still plays an important role in mediating
the coupling between the pump pulse and the excitonic field, thereby
influencing the post-pump spectral response.

\bibliographystyle{apsrev4-2}
\bibliography{biblio}

\end{document}